\documentclass[
  journal=pasa,
  manuscript=research-paper,
  year=2025,
  volume=38,
]{cup-journal}

\usepackage{amsmath}
\usepackage[nopatch]{microtype}
\usepackage{booktabs}
\usepackage{graphicx}	

\title{A Main Sequence CH-star in the globular cluster M55 (NGC~6809)}

\author{G. S. Da Costa}
\affiliation{Research School of Astronomy \& Astrophysics, Australian National University, Canberra, ACT 0200, Australia}
\email[G. S. Da Costa]{gary.dacosta@anu.edu.au}

\author{T. Nordlander}
\affiliation{Research School of Astronomy \& Astrophysics, Australian National University, Canberra, ACT 0200, Australia}
\alsoaffiliation{Department of Physics and Astronomy, Uppsala University, 751 20 Uppsala, Sweden}


\usepackage{amsmath}
\usepackage{hyperref}

\begin{document}

\begin{abstract}
Spectra have been obtained with multi-fibre instrument 2dF on the Anglo-Australian Telescope of 89 candidate main sequence stars in the globular cluster M55 (NGC~6809).  Radial velocities and {\it {Gaia}} proper motions confirm 72 candidates as cluster members.  Among these stars one stands out as having a substantially stronger G-band (CH) than the rest of the member sample.  The star is a dwarf carbon star 
that most likely acquired the high carbon abundance ([C/Fe] $\approx$ 1.2 $\pm$ 0.2) via mass transfer from a $\sim$1$-$3 M$_{\odot}$ binary companion (now a white dwarf) during its AGB phase of evolution.  Interestingly, M55 also contains a CH-star that lies on the cluster red giant branch -- the low central concentration/low density of this cluster presumably allows the survival of binaries that would otherwise be disrupted in denser systems.  The existence of carbon stars in six other globular clusters is consistent with this hypothesis, while the origin of the carbon-enhanced star in M15 (NGC~7078) is attributed to a merger process similar to that proposed for the origin of the carbon-rich R~stars. 
\end{abstract}

\section{Introduction}

Carbon stars are stars with an overabundance of carbon\footnote{Usually taken as [C/Fe] $>$ 0.7 \citep{Aoki2007}} in their atmospheres, and are frequently designated as C-stars, CH-stars, or CEMP (carbon-enhanced metal-poor) stars.
While the existence of such stars has a long history \citep[e.g.][]{Keenan42,Bond74,Dahn77} it is now generally accepted that most examples can be grouped into one of three broad categories.  The first category are stars with masses in the range $\sim$1 -- 5 M$_{\odot}$ in the thermally-pulsing phase on the upper asymptotic giant branch -- the TP-AGB stars \citep[e.g.][]{AKJL07,VA09,Cristallo09,Rosenfield16}.  In TP-AGB stars, carbon is synthesized via triple-$\alpha$ burning in the thermal pulses and is subsequently dredged-up into the surface layers.  These stars also synthesize $s$-process elements, such as Ba, during the thermal pulses and these elements are also dredged-up into the surface layers.
TP-AGB stars are seen in all stellar systems where there is a significant intermediate-age population -- for example in the $\sim$1--3 Gyr old star clusters in the Magellanic Clouds and in the intermediate-age field star populations of the SMC and LMC \citep[e.g.][]{Pasto2020}.

The second category is the so-called CEMP-no stars.  These are carbon-enhanced metal-poor stars with [Fe/H] $<$ --1 \citep[][]{BeersNC2005} that lack any accompanying enhancement of neutron-capture elements, hence the use of ``no'' in the designation.  CEMP-no stars are generally quite metal-poor (e.g. [Fe/H] $\leq$ --3) and become increasingly common as [Fe/H] decreases, dominating ultra-metal-poor star numbers below [Fe/H] = --4 \citep[e.g.][]{Placco14}. The carbon enhancements in these stars are often very large with [C/Fe] values exceeding 2.5 common \citep[e.g.\ Table 7 of][]{FN2015,MSB2015,TN2019}. CEMP-no stars are thought to be predominantly ``second generation'' stars; stars that formed from the enrichment products of zero-metallicity (Population~III) supernovae. Specifically, the abundances reflect the products of low-luminosity mixing-and-fallback Pop~III supernovae \citep[e.g.][]{NKT2013}, where the majority of the supernovae nucleosynthetic products remain in the remnant and only elements from the outer layers, such as C and O, pollute the surrounding interstellar medium \citep[e.g.][]{TN2019,GH2020}.

The third category is where the carbon enhancement results from mass transfer in a binary system. With the appropriate separation the (original) primary can, during its TP-AGB phase, transfer mass to the (original) secondary subsequently evolving to become a white dwarf and leaving the companion carbon-enhanced.  C-stars in this category occur in all stellar populations including the Galactic halo \citep[e.g.][and the references therein]{Lin2024} and the Galaxy's dwarf spheroidal companions \citep[e.g.][]{Kirby2015,Lardo2016, Salgardo2016}.  The classification of metal-poor examples (i.e., CEMP stars ) is discussed in \citet{BeersNC2005}, where the sub-classes CEMP-s, CEMP-r/s, and CEMP-r are defined.  These sub-classes depend on which neutron capture elements are enhanced in the star's atmosphere.  For example, CEMP-s stars, which make up the largest number of CEMP stars \citep{Aoki2007} are defined by [Ba/Fe] $>$ 1.0 and [Ba/Eu] $>$ +0.5 dex. 
The binary nature of at least the large majority of these carbon-enhanced stars has long been established \citep[e.g.][]{McClure1984, McClure1997a,Lucatello2005,Starkenburg2014,Foster2024}.
It is also worth noting that while carbon stars in the first category are by definition giants, carbon stars in the other two categories can, and do, occur as either dwarfs or giants.

As regards carbon stars in globular clusters, an extensive spectroscopic survey is that of $\omega$~Cen (NGC~5139) by \citet{vanLoon2007}.  This study obtained spectra of over 1500 $\omega$~Cen members selected to uniformly sample a ($B, B-V$) colour-magnitude diagram to $B$ magnitudes approximately 3~magnitudes below the red giant branch (RGB) tip.  This survey revealed seven carbon stars, four of which were previously known \citep[see references in][]{vanLoon2007} and three of which were new.  The spectra \citep[see Fig.\ 25 of][]{vanLoon2007} all show C$_{2}$ absorption as well as CN- and CH-bands.  No measurements of [C/Fe] are presented but it is likely that all are significantly enhanced in carbon.  Similarly, the complex nature of the spectra precluded estimates of the barium abundances for these stars but \citet{vanLoon2007} suggested that the carbon enhancements were most likely the result of mass-transfer processes.  The faintest carbon star (LEID~53109) lies near the faint limit of the \citet{vanLoon2007} sample, slightly more than 3 magnitudes below the RGB tip.

Small numbers of CH-stars on the upper part of the RGB have also been identified in six additional clusters.  \citet{GFV2021} have identified 2, and perhaps as many as 5, CH star candidates in M53 (NGC~5024) from spectroscopy of 94 cluster RGB stars.  Similarly, there are at least 3 CH stars in M22 (NGC~6656) \citep{McJEN1977,HesserH79}, 2 in M2 (NGC~7089) \citep{SmithMateo1990, Yong2014} and 1 in each of M55 (NGC~6809) \citep{GHSJEN1982, Briley1993}, NGC~6402 \citep{Cote1997} and NGC~6426 \citep{Sharina2012}. \citet{Kirby2015} has also noted the presence of a CH star with [C/Fe] = 0.9 $\pm$ 0.1 in a sample of $\sim$160 RGB and AGB stars in M15 (NGC~7078).
Overall these numbers are relatively small in the context of the many 100s of globular cluster RGB-star spectra obtained over the past two decades or more \citep[e.g.][]{CarrettaVII,CarrettaVIII}.  An RGB CH-star, which possesses large over-abundances of $s$-process elements, has also been found in the 300S stellar stream \citep{Usman2024}, whose progenitor was a globular cluster. \citet{Usman2024} conclude that the over-abundances in the CH-star result from mass-transfer in a binary system.

In this paper we present the results of a search for dwarf carbon (dC) stars on the main sequence of the metal-poor globular cluster M55.  We also revisit similar data for $\omega$~Cen (NGC~5139), and the globular clusters NGC~6397, NGC~6752 and 47~Tuc (NGC~104) originally presented in \citet{LS2006} and \citet{LS2007}.  The imaging observations used to select the M55 main-sequence sample and the subsequent spectroscopic observations are described in the following section.  The cluster membership and line-strengths of the sample stars are discussed in \S \ref{Sect3} where one M55 member star is shown to be a candidate dC star.  The [C/Fe] ratio is determined via spectral synthesis and confirms the dC designation.  The M55 results and those for the other stellar systems are then discussed in the general context of CH-stars in globular clusters in \S \ref{Sect4}.  Section 5 summarizes the results.

\section{Observations and Reductions}
 
The observations and reductions of both the imaging and spectroscopic data followed the same processes as outlined in \citet{LS2006,LS2007}.  Briefly, M55 was imaged in $B$ and $V$ filters with a Tektronix 2k CCD using the ANU 1m telescope at Siding Spring Observatory.  The CCD camera had a $20^{\prime} \times 20^{\prime}$ field-of-view. Four fields were observed, each centered $\sim$12$^{\prime}$ from the cluster centre and lying to the NE, NW, SE and SW\@.  Exposures were 2 $\times$ 600s in $B$ and 1 $\times$ 500s in $V$ for each field in $\sim2^{\prime\prime}$ seeing under photometric conditions across three nights of observing.  Standardization was provided by observations of \citet{Landolt92} standard star fields each night.

\begin{figure}[t]
	\includegraphics[width=\columnwidth]{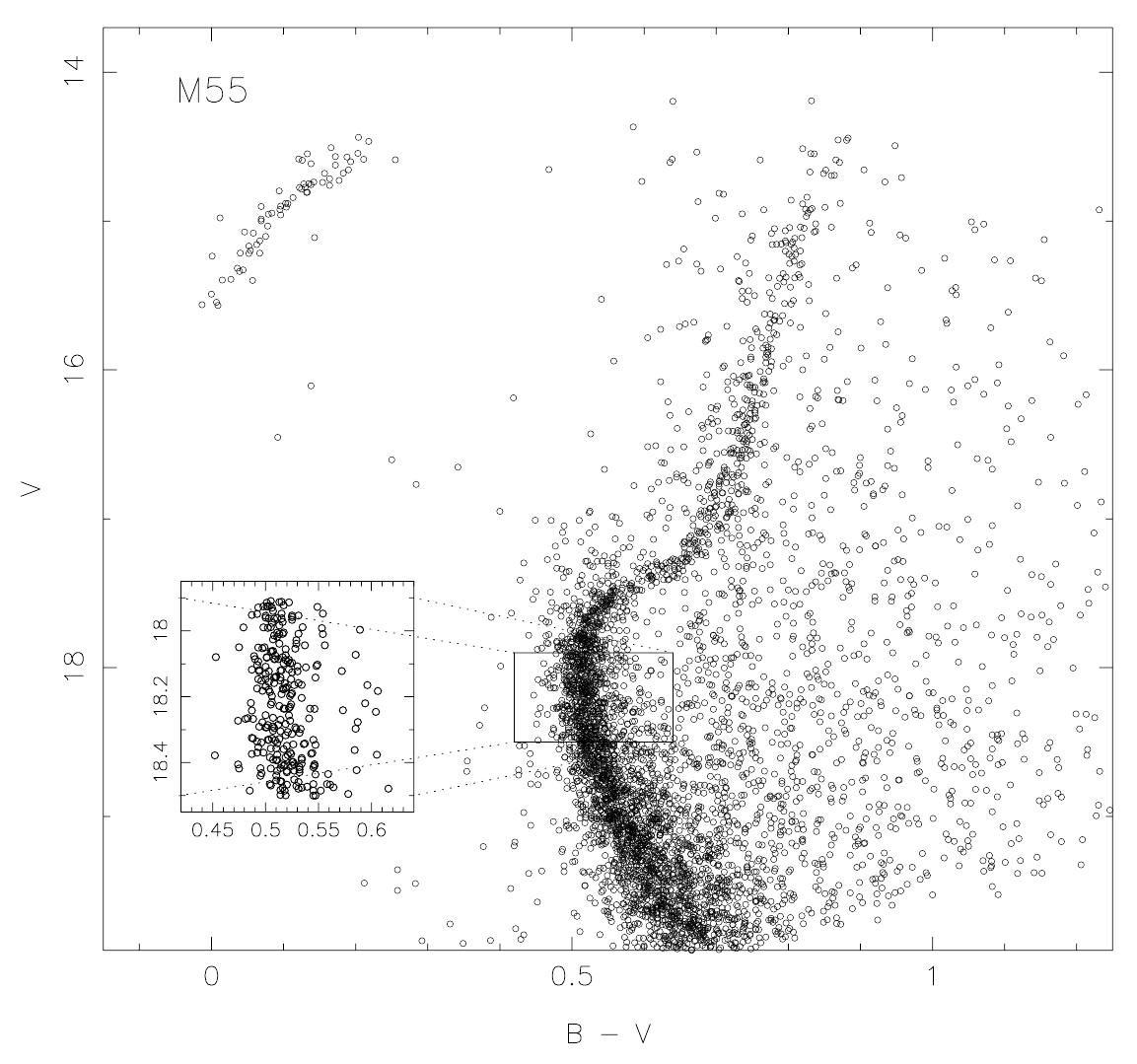}
    \caption{A colour-magnitude diagram for the globular cluster M55 obtained with a Tek 2k CCD on the 1m telescope at Siding Spring Observatory. The insert box shows the selection region for the main-sequence star sample.}
    \label{figure1}
\end{figure}

Stars on the CCD frames were measured using aperture photometry with the stars in common across the overlap regions used to ensure the measured magnitudes were consistent across the fields.  The final dataset consisted of all stars with distance from the cluster centre between 3.6$^{\prime}$ and 13$^{\prime}$, and which had no neighbouring star within 6$^{\prime\prime}$.  The resulting M55 colour-magnitude diagram (CMD) is shown in Fig.\ \ref{figure1}; the bright magnitude limit at $V$ $\approx$ 14.4 is set by saturation on the $V$ frames.  The uncertainty in the photometric zero points is approximately $\pm$0.02 mag for both filters.  The insert box in Fig.\ \ref{figure1}, namely 0.42 $\leq$ $B-V$ $\leq$ 0.64 and 17.9 $\leq$ $V$ $\leq$ 18.5, shows the selection region used to define a main-sequence star sample for potential observation with the Anglo-Australian Telescope's 2dF multi-fibre instrument \citep{Lewis2002}.  A sample of the brightest blue horizontal branch (HB) stars in Fig.\ \ref{figure1} was also selected to serve as candidate fiducial stars for 2dF positioning and guiding.

Positions for the stars in the main sequence sample (and the HB candidate fiducials) were derived using an early version of the USNO astrometric catalogue \citep{Monet96} and the {\sc astrom} basic astrometry code \citep[see][]{Wallace2014}\footnote{As will be evident to the reader, the position determinations took place well before the availability of modern astrometric catalogues such as {\it Gaia} DR3 \citep{GaiaDR3}.}.  With the relatively limited field-of-view of the CCD and the comparative lack of optical distortion in the 1m camera optics, uncertainties in the final positions were of order of 0.3$^{\prime\prime}$, which proved sufficient given the 2dF on-sky projected fibre size of $\sim$2.1$^{\prime\prime}$ \citep{Lewis2002}. Again the stars in the overlap regions between the fields were used to ensure a consistent astrometric solution across the entire field.  The positions of a number of star-free regions were also determined for use as locations for sky fibres.

The main sequence sample shown in the inset box in Fig.\ \ref{figure1} contains 271 stars.  However, because of the concentration towards the cluster centre and the 2dF minimum fibre spacing of 25--30$^{\prime\prime}$ \citep{Lewis2002} only a subsample could be configured for observing.  Running the 2dF fibre allocation code {\sc configure} revealed that only about 25 extra stars could be configured if both spectrographs were employed.  For that reason and because of potential halation issues with spectrograph~2 \citep{Lewis2002}, it was decided to use 
only spectrograph~1.  Eighty nine stars from the full sample were configured along with 40 sky positions.

The spectrograph was configured with the 1200B grating\footnote{Coincidentally \citet{vanLoon2007} in their study of $\omega$~Cen used 2dF with an identical instrumental setup except for a slightly redder central wavelength.} centred at $\lambda$4250\AA\/ giving a wavelength coverage of approximately $\lambda$3700 -- 4750\AA\/ at a resolution of $\sim$2.8\AA.  Four 2000s integrations were obtained under clear skies in 2--2.4$^{\prime\prime}$ seeing.  Also obtained were fibre flats and arc lamp exposures. The data were reduced using the version of the reduction code {\sc{2dfdr}} available at the time.  The individual reduced frames were then co-added.  As a first check, 2.5 $\times$ the logarithm of the average counts in the wavelength region $\lambda$3700 -- 4650\AA\/ for each star observed was plotted against the corresponding $B$ magnitude.  This revealed a relatively small scatter about a unit slope line except for 5 fibres: 3 had virtually no signal indicating a fibre positioning/stellar position mis-match while 2 stars had considerably fewer counts for their $B$ magnitudes compared to the rest of the sample.  These 5 stars were removed from consideration leaving 84 for further analysis.  The S/N ratio in the vicinity of $\lambda$4520\AA\/ for a star with $B$ $\approx$ 18.5 (the median for the observed sample) is approximately 25 per 1.1\AA\/ pixel.

\begin{figure}
	\includegraphics[width=\columnwidth]{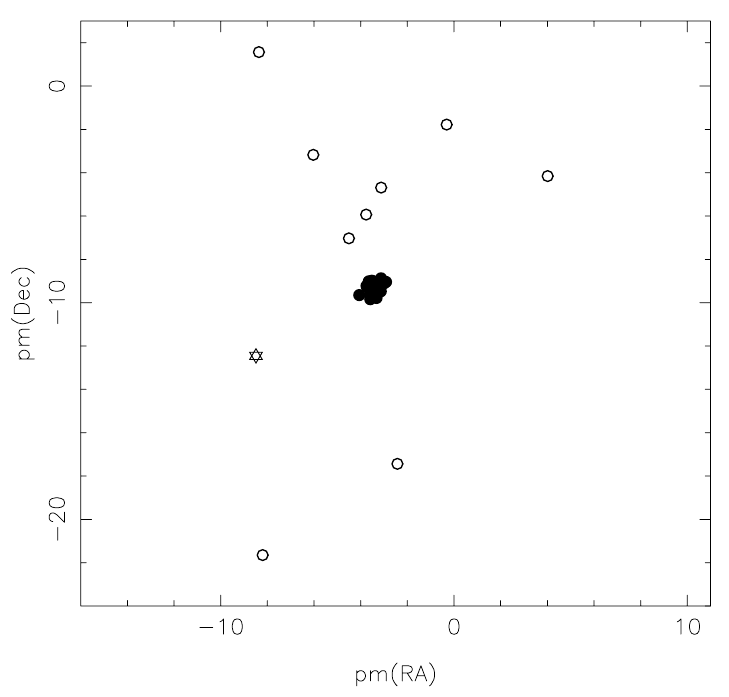}
    \caption{{\it{Gaia}} DR3 proper motions (pm) in mas/yr for the M55 main sequence star sample.  Filled symbols are pm and radial velocity (r$_{\rm v}$) cluster members, open circles are pm and r$_{\rm v}$ non-members, and the open star-symbol is a star that is considered, on the basis of its pm, as a non-member despite having a r$_{\rm v}$ consistent with that of the cluster. }
    \label{figure2}
\end{figure}

\section{Analysis} \label{Sect3}

\subsection{Cluster membership}

As a first pass to determine cluster membership, the 84 individual spectra were cross-correlated with a template spectrum over the wavelength region $\lambda$3700 -- 4500\AA.  The template spectrum consisted of the summed spectra of 3 main sequence stars in the globular cluster NGC~6397, observed with the same instrumental set up during the same observing run.  This revealed 9 stars with clearly discrepant radial velocities compared to the mean velocity for the remainder of the sample.   The 77 remaining stars all lie within $\pm2\sigma$ of the mean with a $\sigma$ value 13 km~s$^{-1}$.  Given that the central velocity dispersion of M55\footnote{See the 4th (March 2023) version of the Baumgardt et al., database {\it ``Fundamental parameters of Galactic globular clusters''}  available at https://people.smp.uq.edu.au/HolgerBaumgardt/globular/} is approximately 5.0 
km~s$^{-1}$ , and that the dispersion decreases with increasing radius, the $\sigma$ value is driven solely by the errors in the individual radial velocities.

Further investigation of the cluster membership status of the observed stars is also possible through the availability of high precision proper motions in the {\it{Gaia}} DR3 data release \citep{GaiaDR3}.  The outcome is plotted in Fig.\ \ref{figure2} which shows the {\it{Gaia}} DR3 proper motions for the 84 candidates.  The nine radial velocity non-members are clearly also proper motion non-members, and a further star is revealed as proper motion non-member despite a radial velocity compatible with membership.  Inspection of the {\it{Gaia}} DR3 database information also reveals that two of the member candidates have probable contributions from multiple stars within the individual fibre field that may bias the observed spectrum.  For example, fibre 136 was positioned on what turned out to be {\it{Gaia}} DR3 sources 6751341388458579328 and 6751341388451408256 that are less than 1$^{\prime\prime}$ apart and which have almost identical $G$ magnitudes.  Removing the two complex objects and the 10 non-members leaves a final sample of 72 M55 main sequence members.  For completeness we note that the mean {\it{Gaia}} DR3 proper motions in RA and Dec for the 72 stars are --3.41 $\pm$ 0.02 mas/yr and --9.32 $\pm$ 0.02 mas/yr (std error of the mean), respectively, values that agree very well with the proper motions for M55 of --3.43 $\pm$ 0.02 mas/yr and --9.31 $\pm$ 0.02 mas/yr given in \citet{VB2021}.

\subsection{Line Strengths}

The strongest features in the observed spectra of the M55 main sequence stars are the H$\gamma$ and H$\delta$ hydrogen lines, the Ca {\sc{ii}} K-line, the blend of the Ca {\sc{ii}} H-line with H$\epsilon$, the G-band of CH and the higher order Balmer lines at the bluest wavelengths.  After shifting the spectra to rest wavelengths using the observed radial velocities, the equivalent widths of the Ca {\sc{ii}} K-line were measured by fitting a gaussian to the line profiles using a feature window of $\lambda$3925--3943\AA\/ and blue and red continuum windows of $\lambda$3910--3920\AA\/ and $\lambda$4015--4060\AA, respectively.  The mean equivalent width for the 72 member stars was 3.35\AA\/ with a standard deviation of 0.47\AA, while the mean error in the equivalent widths, from the gaussian fit parameter uncertainties, was 0.49\AA.  There is therefore no evidence of any instrinsic spread in [Fe/H] for these stars.  
This result is consistent with those of \citet{CarrettaVII, CarrettaVIII} who, based on VLT/GIRAFFE and VLT/UVES spectra of substantial samples of cluster red giants, concluded that for the clusters studied, which included M55, the upper limit on any intrinsic [Fe/H] dispersion is $\sim$0.05 dex. \citet{Rain2019} reached a similar conclusion based on high-resolution spectra of 11 M55 RGB stars.

On the other hand, \citet{Milone17} argued, based on HST photometry, that the colour spread for the first population (1P) stars in their globular cluster chromosome diagrams exceeds that expected from the photometric errors, suggesting that the 1P stars are not a chemically homogeneous population.  M55 is included in the \citet{Milone17} sample.  In the follow-up study of \citet{Legnardi22} \citep[see also][]{Latour25}, RGB isochrones from \citet{Dotter08} were used to infer $\delta[Fe/H]_{1G}$ values from the \citet{Milone17} chromosome photometry, where $\delta[Fe/H]_{1G}$ is the 10--90th percentile range of the derived [Fe/H] values.  For M55, $\delta[Fe/H]_{1G}$ = 0.115 $\pm$ 0.011 which, assuming a gaussian metallicity distribution corresponds to $\sigma$[Fe/H] $\approx$ 0.04 dex, consistent with the \citet{CarrettaVII,CarrettaVIII} upper limit.

\begin{figure}[t]
	\includegraphics[width=\columnwidth]{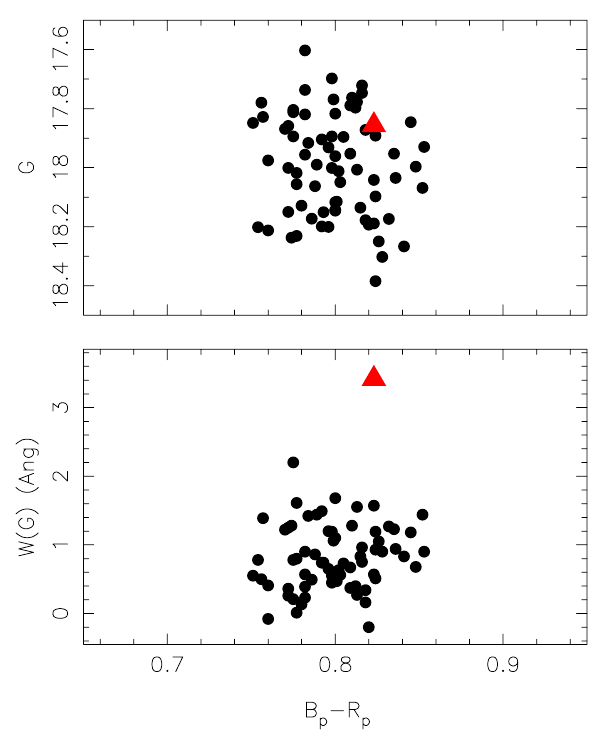}
    \caption{{\it{Upper panel}}. {\it{Gaia}} DR3 ($G$, $B_{p}-R_{p}$) colour-magnitude diagram for the 72 M55 main sequence member stars. {\it{Lower panel}}. Equivalent width of the G-band feature of CH (W(G)) in Angstroms plotted against ($B_{p}-R_{p}$) colour.  In both panels the red triangle is star 307649 ({\it Gaia} DR3 6751343892420411136.)}.   
    \label{figure3}
\end{figure}

The lack of any substantial intrinsic metallicity variations in M55 is also attested by the CMD for the main sequence member stars studied here.  The CMD, based on {\it Gaia} DR3 photometry, is shown in the upper panel of Fig.\ \ref{figure3}.  Here the ($B_{p} - R_{p}$)  colour distribution has a width characterized by $\sigma$($B_{p} - R_{p}$) = 0.025 mag.  This value is consistent with that expected given {\it {Gaia}} photometry errors of $\sim$0.010 -- 0.015 mag \citep{GaiaDR3} together with reddening variations across the observed cluster field of order 0.02 mag in E($B-V$) \citep{AlonsoG2012}, using the $B_{p}$ and $R_{p}$ extinction corrections of either \citet{Luca2021} or \citet{ZhangYuan2023}.

\begin{figure*}
	\includegraphics[width=\textwidth]{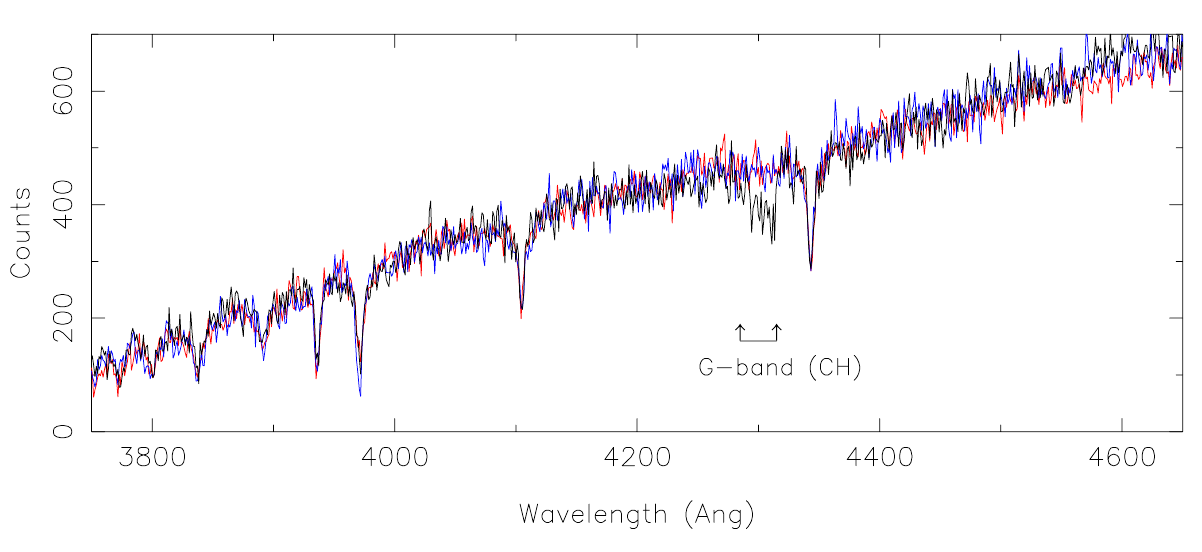}
    \caption{Comparison of the observed spectrum of M55 star 307649, plotted in black, with those of M55 stars 305079 and 202495 plotted in red and blue, respectively. All three stars have very similar magnitudes and colours.  Note that the only significant difference between the three spectra is the much stronger G-band in the spectrum of star 307649.  The comparison star spectra have been normalized to that of 307649 in the vicinity of 4500\AA.}
    \label{figure4}
\end{figure*}

The equivalent width of the G-band feature of CH in the spectra of the main sequence stars, W(G),  was determined by numerical integration {\bf (rectangular rule)}.  The feature window was $\lambda$4295--4315\AA\/ with blue and red continuum windows of $\lambda$4255--4280\AA\/ and $\lambda$4470--4520\AA, respectively.  The results are shown in the lower panel of Fig.\ \ref{figure3}.  Here one star, with ID 307649 ({\it Gaia} DR3 6751343892420411136), has a substantially stronger G-band than the remainder of the sample.  The mean value of W(G) for the other 71 stars is 0.81\AA\/ with a standard deviation of 0.47\AA; the value for 307649 is larger than the mean by 5.5$\sigma$ and larger than the next highest W(G) value by 2.2$\sigma$.  

The difference is illustrated in Fig.\ \ref{figure4} where the observed spectrum of 307649 is compared with those for stars 305079 ({\it Gaia} DR3 6751392374007714176) and 202495 \\
(6751393271659171200); all three stars have closely similar $G$ and ($B_{p}-R_{p}$) values.  The spectra of the comparison stars have been normalized to that of 307649 in the vicinity of 4500\AA.  It is immediately evident from Fig.\ \ref{figure4} that apart from the much stronger G-band in 307649, the three spectra are otherwise very similar.  Star 307649 is thus likely to be a carbon-enhanced main sequence star member of the globular cluster M55.  Unfortunately, as is evident from Fig.\ \ref{figure4}, the S/N of the spectra is insufficient to draw any conclusions regarding any possible enhancement in nitrogen in this star -- neither of the CN-bands at $\lambda$3883\AA\/ or $\lambda$4215\AA\/ are obviously present, although all three stars are likely too hot for significant formation of CN in the atmospheres.  Similarly, there is no evidence for any possible increased line-strength for the $s$-process elements Ba (via $\lambda$4554\AA\/ Ba {\sc{ii}}) and Sr (via $\lambda$4077\AA\/ Sr {\sc{ii}}) in the spectrum of 307649 relative to those for stars 305079 and 202495. Higher resolution and/or higher S/N is required to investigate the N, Sr and Ba abundances as well as those of other elements.

\subsection{The [C/Fe] value for M55 star 307649} 

The effective temperature (T$_{\rm eff}$) of the star was estimated from the {\it{Gaia}} DR3 photometry \citep{GaiaDR3} via the \mbox{T$_{\rm eff}$ -- colour} relations of \citet{Luca2021}.  A reddening of E($B-V$) = 0.12 mag \citep{AlonsoG2012} was assumed, as well as log~$g$ $\approx$ 4.2 and a metal abundance [Fe/H] = --1.9 \citep{Carretta2009}.  The resulting value is T$_{\rm eff}$ = 6100 K with a conservative uncertainty of $\pm$100 K. Given that the distance to M55 is well-established as 5.3 $\pm$ 0.1 kpc \citep{BV2021}, the surface gravity of 307649 
is log $g$ = 4.1 in cgs units for the assumed temperature, using a bolometric correction to the $G$ mag from \citet{LC2018} and an assumed mass of 0.8 M$_{\odot}$. 

Synthetic spectra were generated following the approach outlined in \citet{TN2019}. In brief, plane-parallel MARCS model atmospheres \citep{Gustafsson2008} were employed together with the {\tt{TurboSpectrum}} code \citep[v15.1;][]{Plez2012}. A microturbulence velocity $v_{\rm mic}$ of 1~km~s$^{-1}$ was adopted and the spectra calculated assuming [Fe/H] = --1.9 \citep{Carretta2009}, and an enhancement [$\alpha$/Fe] = 0.4 (for O, Ne, Mg, \dots, Ti), with the reference solar composition that of \citet{Asplund09}.  The assumed [O/Fe] value is somewhat higher than the observed mean value of [O/Fe] = 0.16 for 14 M55 red giants\footnote{The three M55 red giants with the lowest [Na/Fe] values in \citet{CarrettaVIII} have $\langle$[O/Fe]$\rangle$ $\approx$ 0.26 dex.} \citep{CarrettaVIII}, but the difference has a negligible effect on the derived [C/Fe] value.  Spectra were computed for a range of [C/Fe] values, using the CH linelist of \citet{Masseron2014}, in addition to atomic lines from VALD3 \citep{VLAD}.  The best-fit of the synthetic spectra was determined via a $\chi^2$ minimization process in which synthetic spectra with different [C/Fe] values were compared with the continuum normalized observed spectrum over the wavelength interval $\lambda$4150--4450\AA.

\begin{figure}
	\includegraphics[width=\columnwidth]{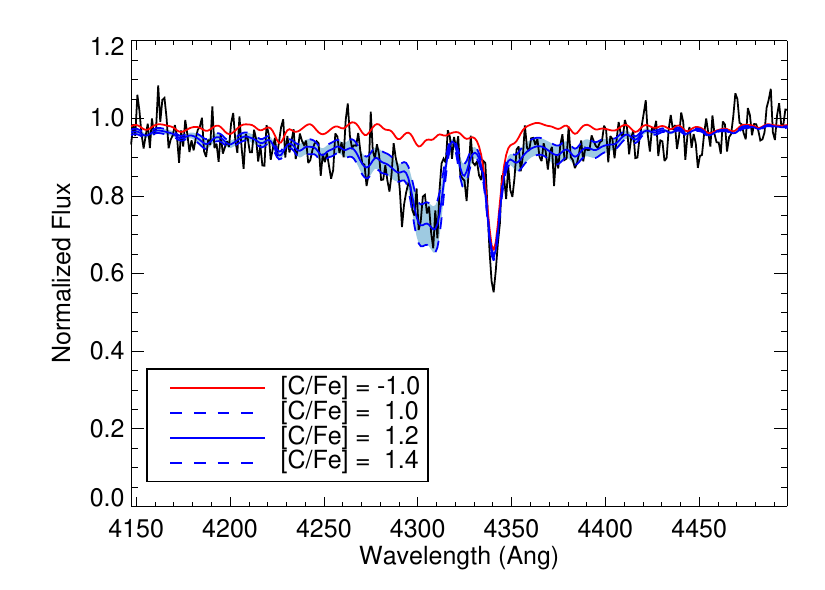}
    \caption{Synthetic spectrum fits to the observed spectrum of star M55 307649.  The solid blue line is the best fit ([C/Fe] = +1.2) for the adopted parameters, while the dashed blue lines have [C/Fe] $\pm$0.2 about this value.  The region between the dashed lines is shaded in light blue to emphasize the regions sensitive to [C/Fe].  The red line shows the synthetic spectrum with the same parameters except that [C/Fe] = --1.0 dex.}  
    \label{figure5}
\end{figure}

Synthetic spectra fits are shown in Fig.\ \ref{figure5}.  The best-fit is for [C/Fe] = 1.2 $\pm$ 0.2, where the uncertainty includes that from the $\pm$100~K temperature uncertainty.  With this level of carbon-enhancement, M55 307649  ({\it{Gaia}} DR3 6751343892420411136) can definitely be classified as a dwarf carbon (dC) star.  Unfortunately however, the temperature, combined with the resolution and signal-to-noise of the observed spectrum, preclude any statement regarding potential enhancements in $s$-process elements, such as Sr or Ba, in the atmosphere of the star.  Further classification, e.g., as CEMP-s, must await higher quality data. Specifically, spectrum synthesis calculations for the 
$\lambda$4554\AA\/ Ba{\sc ii} line show that, for this combination of metallicity, effective temperature and spectral resolution, a S/N in excess of 200 would be required to distinguish [Ba/Fe] = +1.0 from [Ba/Fe] = 0.0 at the 3$\sigma$ level.  On the other hand, at higher resolution, e.g., R $\approx$ 5400 with the UVB arm of the ESO VLT/X-shooter instrument \citep{vernet11}, spectra with S/N $>$ $\sim$50 can readily measure a [Ba/Fe] abundance excess of $\sim$1 dex or higher in this star relative to similar temperature M55 main sequence stars.
\\
\\
\\

\section{Discussion} \label{Sect4}

\subsection{Other samples}

As discussed in \citet{LS2006,LS2007} main sequence stars in the globular clusters $\omega$~Cen, NGC~6397, NGC~6752 and 47~Tuc were observed with the same instrument configuration as that described here, except that for 47~Tuc and $\omega$~Cen both 2dF spectrographs were used.  The main sequence stars were selected from similar ($V, B-V$) CMDs in the same way as for the M55 sample and occupy a similar M$_{V}$ range.  For NGC~6397 and NGC~6752 approximately 120 main sequence star members were analyzed while for 47~Tuc the sample size was approximately 280 members. As demonstrated in the right panels of Fig.\ 6 of \citet{LS2007}, there are no main sequence stars in these three cluster samples with significantly enhanced [C/Fe] values, although the dichotomy of [C/Fe] values in 47 Tuc as result of the abundance anomalies in this cluster is evident \citep[results from the 47~Tuc main sequence sample are discussed in more detail in][]{GDaC2004}.

For $\omega$~Cen, as described in \citet{LS2006}, two samples of stars have been observed.  The first is an unbiased main sequence sample, and the second a sample with luminosities above the main sequence turnoff across a range of metallicities.  There are a total of 442 cluster members in both samples with the numbers split approximately equally between the two \citep[see Fig.\ 3 of][]{LS2006}.  \citet{LS2007} divide the set of observed cluster members into a main sequence group ($V$ $>$ 18, M$_{V} > 4.04$) and a subgiant sub-group ($V$ $\leq$ 18, M$_{V} \leq 4.04$).  Each group is then further split by metallicity with stars with [Fe/H] $<$ --1.5, --1.5 $\leq$ [Fe/H] $<$ --1.1, and [Fe/H] $\geq$ --1.1 considered separately. 

Of particular interest here are the left panels of Fig.\ 12 of \citet{LS2007} which show, as a function of M$_{V}$, [C/Fe] values derived from spectrum synthesis, as well as regions where the abundance ratio cannot be reliably determined.  The stars analyzed are drawn from the left panels of Fig.\ 6 of \citet{LS2007} and the typical error in the [C/Fe] determinations is $\pm$0.27 dex \citep{LS2007}.  For the most metal-poor population, there are two main sequence stars with [C/Fe] $\approx$ 1.3 and two subgiants with [C/Fe] values of 1.0 and 0.75, approximately.  For the intermediate metallicity group there is one subgiant with [C/Fe] $\approx$ 0.9 dex.  All five of these stars qualify as C-stars with the two $\omega$~Cen main sequence stars being dC stars\footnote{We note that the \citet{LS2007} Fig.\ 12 [C/Fe] panel for the most metal-rich population shows one subgiant apparently with [C/Fe] $\approx$ 1.0.  However, there is no corresponding entry in the on-line version of Table 3, so we suspect the plotted point is erroneous.}.

\begin{figure}[t]
	\includegraphics[width=\columnwidth]{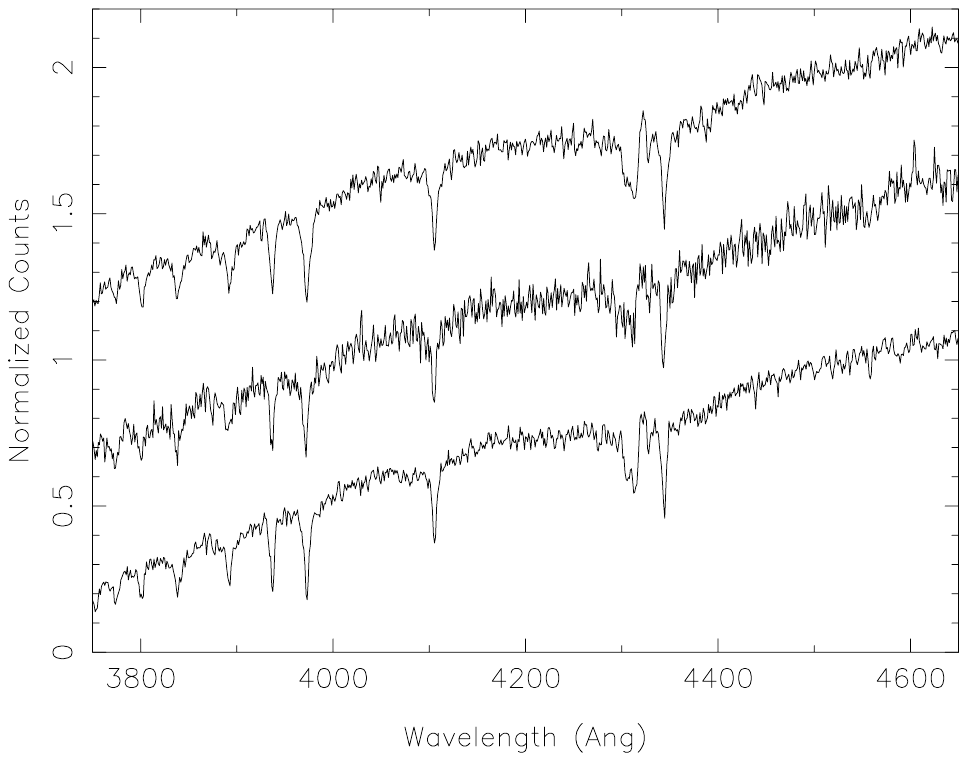}
    \caption{The observed spectrum of star 307649 in M55 (middle) is compared the observed spectra of $\omega$~Cen main sequence stars 7007334 (bottom) and 9005309 (top) from \citet{LS2007}.  The spectra have been normalized at $\sim \lambda$4500\AA.  The middle and upper spectra have been offset vertically by 0.5 and 1.0, respectively.  The $\omega$~Cen spectra have S/N $\approx$ 45 per pix while S/N $\approx$ 25 for the M55 star.  Note the strong G-band of CH in all three spectra.
    }  
    \label{figure6}
\end{figure}

The properties of the M55 dC star and those for the \citet{LS2007} $\omega$~Cen C-stars are given in Table \ref{tab1}.  Listed are the cluster and star IDs, the corresponding {\it Gaia} DR3 IDs, and the T$_{\rm eff}$, log~{\it g}, [Fe/H] and [C/Fe] values, with those for the $\omega$~Cen stars taken from the on-line version of Table~3 of \citet{LS2007}. We note that $\omega$~Cen star 7011049 has a separate second entry in the Table as star 8001811.  The atmospheric parameters are essentially identical for what are presumably separately analyzed spectra.  For example, the two [C/Fe] values are 0.80 and 0.70, respectively; the average values are given in Table \ref{tab1}.  Figure \ref{figure6} shows the observed spectra for the M55 dC star and for the two $\omega$~Cen dC stars: all three spectra are remarkably similar.

\begin{table*}
\centering
\caption{Properties of main sequence and subgiant carbon stars in M55 and $\omega$~Cen.}
\label{tab1}
\begin{tabular}{lrcccll} 
		\hline
	Cluster & ID & {\it Gaia} DR3 ID & T$_{\rm eff}$ (K) & log~{\it g}  (cgs) & [Fe/H] & [C/Fe] \\
	\hline
M55 & 307649 & 6751343892420411136 & 6100 & 4.1 & --1.9 & 1.2 \\
$\omega$ Cen & 7007334 & 6083727681943908992 & 6152 & 4.3 &--1.84  & 1.35\\
$\omega$ Cen & 9005309 & 6083497441641348352 & 6107 & 4.3 & --1.88 & 1.30 \\
$\omega$ Cen & 2010136 & 6083676043580595712 & 5631 & 3.8 & --1.75 & 1.00 \\
$\omega$ Cen & 7011049 & 6083725624687591680 & 5910 & 3.9 & --1.54 & 0.75 \\
$\omega$ Cen & 9004288 & 6083499258399977088 & 5824 & 3.9 & --1.30 & 0.90 \\
	\hline
\end{tabular}
\end{table*}

Two other studies deserve mention.  The first is the extensive survey of stars in the globular cluster M4 (NGC~6121) by \citet{Malavolta}.  These authors analyzed medium-resolution spectra with varying S/N,  obtained with the\\
FLAMES+GIRAFFE multi-object spectrograph at the VLT, for 322 RGB stars and 1869 subgiant and main sequence stars in the cluster.  The spectra covered the narrow wavelength interval 5143\AA\/ $<$ $\lambda$ $<$ 5356\AA, a region that includes a Swan band of C$_{2}$ with bandhead at 5163\AA. While it is likely that the subgiant and main sequence stars are too hot for significant C$_{2}$ band formation even at high [C/Fe] values, this is unlikely to be the case for the M4 RGB stars.  For example, \citet{Kirby2015} notes that the C$_{2}$ bandhead at 5163\AA\/ is weakly present in the M15 CH star that has [C/Fe] = +0.9 $\pm$ 0.1 \citep{Kirby2015} and which lies well below the RGB tip in the M15 CMD \citep{Kirby2015}.  \citet{Malavolta}, however, make no mention of the detection of the C$_{2}$ bandhead in any of their spectra, from which it is reasonable to assume there are no carbon enhanced stars in their M4 RGB sample. 

The second study is that of \citet{DOrazi2010}.  In this work [Ba/Fe] values were determined for 1205 RGB stars across 15 globular clusters using FLAMES/Giraffe high-resolution spectra \citep{CarrettaVII}.  The intention was to identify the presence of any barium-stars in the sample.  Ba-stars generally possess spectra enhanced in carbon and $s$-process elements such as Ba, with the enhancements the results of mass transfer in a binary system.  For example, \citet{McClure1984} and \citet{McClureWood} have shown that all the Ba-stars studied are long period, single line spectroscopic binaries as expected for the mass-transfer origin of the abundance enhancements.  It is worth noting, however, that the spectra analyzed by \citet{DOrazi2010} do not cover any spectral features indicative of an enhanced carbon abundance in any Ba-stars discovered.

\citet{DOrazi2010} define ``Ba-stars'' as any star in a globular cluster where the [Ba/Fe] abundance ratio exceeds by 3.5$\sigma$ the mean [Ba/Fe] value for the cluster.  With this criterion, 5 Ba-stars were identified in 5 globular clusters.  Additional information is now available for 3 of these 5 stars, but not for star 200095 in NGC~288, which also has enhanced [La/Fe] and [Nd/Fe] \citep[see][]{DOrazi2010}, and star 608024 in NGC~6397.  Star 30952 in 47~Tuc is neither enhanced in Ba or carbon: \citet{Dobrov} give [Ba/Fe] = 0.13 which is close to cluster mean, as distinct from the \citet{DOrazi2010} value of 0.51 $\pm$ 0.15 dex.  Similarly, the SDSS DR17 release \citep{Abdurro} lists [C/Fe] = 0.06 for this star.  Star 28903 in NGC~6254 is also not C-enhanced: the SDSS DR17 value is [C/Fe] = --0.39, while \citet{Gerber2018} give [C/Fe] = --0.36 dex.  Finally, for star 38291 in NGC~6752, SDSS DR17 lists [C/Fe] = --0.16, so again the star in not carbon enhanced. The updated results of \citet{DOrazi2010} then suggest that the occurrence of CH-stars in globular clusters is generally a rare event.  

As an aside we note that stars such as 28903 in NGC 6254, which is clearly enhanced in Ba \citep[see Fig.\ 3 of][]{DOrazi2010}, are apparently similar to a small number of globular cluster red giants that are $s$-process enhanced but not carbon-enhanced, i.e., they do not show the characteristics expected from AGB star mass-transfer in a binary system.  Other examples include star 61005163 in NGC 5824 \citep{Roderer16} which has large excesses of Ba, La, Ce and Nd relative to the other cluster red giants, while the r-process elements Sm, Eu, Dy are not distinguished.  \citet{Roderer16} give an evolutionary mixing corrected estimate of [C/Fe] $\approx$ 0.4 for this star, so it is not carbon-enhanced.  Star 4710 in 47 Tuc \citep{Cordero15}, is similar: [Ba/Fe] = 1.0 $\pm$ 0.2, [La/Fe] = 1.1 $\pm$ 0.14 and [Ce/Fe] = 1.0 $\pm$ 0.2 \citep{Cordero15} but with [C/Fe] = 0.15 $\pm$ 0.02 (and [Ce/Fe] = 0.84 $\pm$ 0.05) from the SDSS DR17 release \citep{Abdurro}.  So again the star is not carbon-enhanced.  As discussed in \citet{Roderer16}, the origin of the abundances in these equally rare stars is not readily understood.

\subsection{The origin of CH stars in globular clusters}

The most likely explanation for the presence of carbon enhanced stars in globular clusters is that they result from mass transfer in a binary system that occurs when the (original) primary undergoes thermal pulse driven dredge-up on the upper asymptotic giant branch.  Consequently, the binary must survive in the dynamic environment of a globular cluster long enough for the original primary to reach the TP-AGB phase.  For a $\sim$1.5 M$_{\odot}$ primary, this maybe as long as a few Gyr.  At the same time the binary separation must remain such that the large radius of the TP-AGB star can be accommodated with stable mass transfer, avoiding a common-envelope phase.  

The dynamic environment of globular clusters is capable of strongly influencing binary star members.  Stellar encounters within a cluster cause  ``soft'' binaries (i.e., those with relatively large separations) to disrupt while ``hard'' binaries (i.e., those with smaller separations) become more compact \citep[e.g.,][]{Hut1992}.  In the context of binary CH stars, \citet{Cote1997} give an equation for the characteristic final separation of surviving binaries, $a_{h}$, that depends on the ratio of the central 1D velocity dispersion to the central mass density of a cluster.  They then demonstrate that for M14 and $\omega$~Cen, the values of $a_{h}$ exceed the semi-major axis of the shortest period binary in the \citet{McClureWood} sample of field CH stars.  Thus the existence of mass-transfer binary CH stars in these clusters is plausible.  \citet{GFV2021} reach a similar conclusion regarding the CH stars in M53 (NGC~5024).  As noted by \cite{Cote1997} and \citet{GFV2021}, the equation for $a_{h}$ strictly applies only at the centre of the cluster and does not encompass the complexity of the dynamical evolution of binaries in a globular cluster.

\begin{figure}
	\includegraphics[width=\columnwidth]{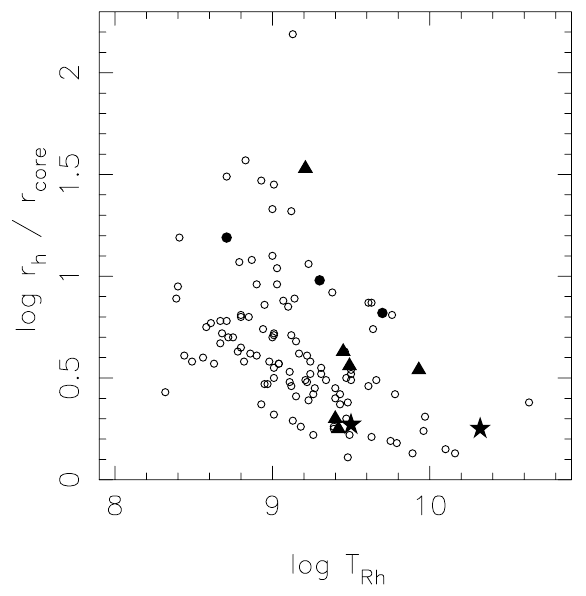}
    \caption{A plot of the logarithm of the ratio of the projected half-mass radius (r$_{\rm h}$) to the core radius (r$_{\rm core}$) against the logarithm of the half-mass relaxation time (T$_{\rm Rh}$) for clusters with masses exceeding $5 \times 10^{4}$ solar masses. Values are taken from the Baumgardt et al.\ globular cluster database (edition 4). M55 and $\omega$~Cen are plotted as star symbols, while NGC~6402, NGC~6426, M53, M22, M2 and M15 (in order of increasing log(r$_{\rm h}$/r$_{\rm core}$) are plotted as filled triangles.  The filled circles are 47~Tuc, NGC~6752 and NGC~6397, also in order of increasing log(r$_{\rm h}$/r$_{\rm core}$).
    }  
    \label{figure7}
\end{figure}

Here, noting that \citet{McJEN1977} were the first to point out that there is a tendency for CH stars to occur in low central concentration clusters, we have adopted a different approach.  Fig.\ \ref{figure7} shows a plot of the logarithm of the ratio of the cluster (projected) half-mass radius, r$_{\rm h}$, to the cluster core radius, r$_{\rm core}$, against the logarithm of the half-mass relaxation time, T$_{\rm Rh}$, in Gyr.  The values are taken from the Baumgardt et al.\ globular cluster database\footnote{Edition 4; https://people.smp.uq.edu.au/HolgerBaumgardt/globular/}; only clusters with masses exceeding $5 \times 10^{4}$ M$_{\odot}$ are plotted: with a mass of 7.3 $\pm$ 2.4 $\times 10^{4}$ M$_{\odot}$ NGC~6426 is the lowest mass cluster containing a CH star. The r$_{\rm h}$ to r$_{\rm core}$ ratio is a measure of the degree of central concentration of a cluster (less centrally concentrated clusters have smaller values) while T$_{\rm Rh}$ is a measure of the extent of dynamical evolution in a cluster.

In the figure the filled 5-point star symbols are M55 and $\omega$~Cen, the two clusters with identified populations of both dwarf and red giant CH stars.  The other six clusters with CH-star RGB members are plotted as filled triangles.  The three filled circles are 47~Tuc, NGC~6752 and NGC~6937 which, based on existing data, do not contain either dwarf or giant CH stars.  The cluster M4 is not separately identified but lies at (8.94, 0.74) in the figure.  With the exception of M15 (see below), the clusters with CH stars show a preference for longer T$_{\rm Rh}$ and lower r$_{\rm h}$ to r$_{\rm core}$ values relative to the overall population.  This is consistent with the interpretation that these clusters have allowed the survival of binaries capable of mass transfer during the original primary's TP-AGB phase.  Note that it is not required that the current CH stars remain in a binary system --- the binary could have been disrupted dynamically subsequent to the mass-transfer event.

Compared to the points in Fig.\ \ref{figure7} for the other 7 globular clusters containing CH stars, the location of M15 in this diagram is anomalous.  Quantitatively, excluding M15, the mean value of $log(r_{h}/r_{core})$ for the clusters with CH-star members is 0.40 $\pm$ 0.06 (std error of mean), while that for M15, at 1.53, is substantially higher.  This discrepancy suggests that the carbon star in M15 may have been produced by a mechanism different from mass transfer in a binary system.  Although on the AGB in the M15 CMD \citep{Kirby2015}, the star is well below the RGB-tip. It therefore lacks the required luminosity to be in the TP-AGB phase, even if it had somehow acquired sufficient mass to be capable of evolving to this phase.  \citet{Kirby2015} suggested that the star could be the product of a merger, which are more likely to occur in high central density, high central concentration clusters like M15 \citep[e.g.][]{Kremer2020}.  The merger process though would be required to produce temperatures hot enough for triple-$\alpha$ burning to generate a substantial carbon overabundance in the merger product.

This question is addressed in studies of the origins of early-type R stars: carbon-rich red giants that are not binaries \citep{McClure1997b}, generally lack enhancements in $s$-process elements, and which are not luminous enough to be in the TP-AGB phase of evolution.  Conventional single-star stellar evolutionary models cannot explain these stars, but as suggested by \citet{McClure1997b} they may have their origin in a merger event.   \citet{Izzard2007} investigated this possibility by calculating binary star evolution models \citep[see also][]{Zhang2020}, and found that merger of a helium white dwarf (HeWD) with a red giant during a common envelope phase is followed by a helium flash in a rotating core that allows the newly generated carbon to be mixed into the surface layers, creating a single star with the characteristics of R stars \citep[see also][]{Zhang2020}.  

Such a binary merger process would then seem to be a plausible explanation for the origin of the M15 CH star: confirmation would follow if the star is not (now) a binary and if it lacks any enhancement in $s$-process elements.  Assuming the merger explanation is correct, the M15 star is then  ``the exception that proves the rule'' \footnote{A phrase attributed originally to the Roman statesman Cicero in his 59 BCE defence of Lucius Balbus (see https://en.wikipedia.org/wiki/Exception\_that\_proves\_the\_rule).} --- the ``rule'' in this case being that CH-stars in low central concentration globular clusters are mass-transfer binaries.


\section{Conclusions}

In this paper we have discussed 2dF spectroscopy of 72 main sequence member stars in the globular cluster M55.  One star was found to have a substantially stronger G-band (CH) compared to the other cluster members.  Spectrum synthesis yielded [C/Fe] $\approx$ 1.2 $\pm$ 0.2 for this star, allowing its classification as a dwarf carbon star.  M55 then joins $\omega$~Cen (NGC~5139) as a globular cluster in which both dwarf and red giant CH-stars are known members.  A compilation of the properties of the 8 Galactic globular clusters containing at least one CH-star member shows that 7 of the clusters have relatively low central concentrations and long half-mass relaxation times.  This is consistent with the origin of the CH-stars as a consequence of mass transfer in binary systems.  The remaining CH-star is found in the high central concentration cluster M15 (NGC~7078) and its origin is postulated as resulting from a HeWD+RGB binary common-envelope merger process, as discussed by \citet{Izzard2007} and \citet{Zhang2020}.  While carbon stars are currently relatively rare objects in globular cluster stellar populations, it is likely that additional examples will be discovered in forthcoming large scale spectroscopic surveys such as 4MOST \citep{deJong2019}, WEAVE \citep{Shoko2024}, and the SDSS-V Milky Way Mapper program (see https://www.sdss.org/dr18/mwm/about/), allowing further investigations of the origins of these stars.

\paragraph{Acknowledgments}

We thank the anonymous referee for their helpful comments. The original project that resulted in these observations was conceived in collaboration with Sean Ryan (Univ.\ of Hertfordshire) and Tad Pryor (Rutgers University).  Russell Cannon (AAO) provided invaluable support for the 2dF observations for which we remain grateful.

This work has made use of data from the European Space Agency (ESA) mission
{\it Gaia} (\url{https://www.cosmos.esa.int/gaia}), processed by the {\it Gaia}
Data Processing and Analysis Consortium (DPAC,
\url{https://www.cosmos.esa.int/web/gaia/dpac/consortium}). Funding for the DPAC
has been provided by national institutions, in particular the institutions
participating in the {\it Gaia} Multilateral Agreement.

This research has made use of the VizieR catalogue access tool, CDS,
 Strasbourg, France \citep{10.26093/cds/vizier}. The original description 
 of the VizieR service was published in \citet{vizier2000}.

The authors acknowledge the traditional owners of the land on which Siding Spring Observatory is located, the Gamilaraay people, and pay our respects to elders past, present and emerging.

\paragraph{Data Availability Statement}

The raw data from the 2dF observations are available in the AAT data archive at \\
https://datacentral.org.au/. The reduced spectra and the 1m photometry will be shared on reasonable request to the first author.





\bibliography{references_rev.bib}

\begin{thebibliography}{}
\expandafter\ifx\csname natexlab\endcsname\relax\def\natexlab#1{#1}\fi

\bibitem[{{Abdurro'uf} {et~al.}(2022){Abdurro'uf}, {Accetta}, {Aerts}, {Silva Aguirre}, {Ahumada}, {Ajgaonkar}, {Filiz Ak}, {Alam}, {Allende Prieto}, {Almeida}, {Anders}, {Anderson}, {Andrews}, {Anguiano}, {Aquino-Ort{\'\i}z}, {Arag{\'o}n-Salamanca}, {Argudo-Fern{\'a}ndez}, {Ata}, {Aubert}, {Avila-Reese}, {Badenes}, {Barb{\'a}}, {Barger}, {Barrera-Ballesteros}, {Beaton}, {Beers}, {Belfiore}, {Bender}, {Bernardi}, {Bershady}, {Beutler}, {Bidin}, {Bird}, {Bizyaev}, {Blanc}, {Blanton}, {Boardman}, {Bolton}, {Boquien}, {Borissova}, {Bovy}, {Brandt}, {Brown}, {Brownstein}, {Brusa}, {Buchner}, {Bundy}, {Burchett}, {Bureau}, {Burgasser}, {Cabang}, {Campbell}, {Cappellari}, {Carlberg}, {Wanderley}, {Carrera}, {Cash}, {Chen}, {Chen}, {Cherinka}, {Chiappini}, {Choi}, {Chojnowski}, {Chung}, {Clerc}, {Cohen}, {Comerford}, {Comparat}, {da Costa}, {Covey}, {Crane}, {Cruz-Gonzalez}, {Culhane}, {Cunha}, {Dai}, {Damke}, {Darling}, {Davidson}, {Davies}, {Dawson}, {De Lee}, {Diamond-Stanic}, {Cano-D{\'\i}az}, {S{\'a}nchez},
  {Donor}, {Duckworth}, {Dwelly}, {Eisenstein}, {Elsworth}, {Emsellem}, {Eracleous}, {Escoffier}, {Fan}, {Farr}, {Feng}, {Fern{\'a}ndez-Trincado}, {Feuillet}, {Filipp}, {Fillingham}, {Frinchaboy}, {Fromenteau}, {Galbany}, {Garc{\'\i}a}, {Garc{\'\i}a-Hern{\'a}ndez}, {Ge}, {Geisler}, {Gelfand}, {G{\'e}ron}, {Gibson}, {Goddy}, {Godoy-Rivera}, {Grabowski}, {Green}, {Greener}, {Grier}, {Griffith}, {Guo}, {Guy}, {Hadjara}, {Harding}, {Hasselquist}, {Hayes}, {Hearty}, {Hern{\'a}ndez}, {Hill}, {Hogg}, {Holtzman}, {Horta}, {Hsieh}, {Hsu}, {Hsu}, {Huber}, {Huertas-Company}, {Hutchinson}, {Hwang}, {Ibarra-Medel}, {Chitham}, {Ilha}, {Imig}, {Jaekle}, {Jayasinghe}, {Ji}, {Johnson}, {Jones}, {J{\"o}nsson}, {Katkov}, {Khalatyan}, {Kinemuchi}, {Kisku}, {Knapen}, {Kneib}, {Kollmeier}, {Kong}, {Kounkel}, {Kreckel}, {Krishnarao}, {Lacerna}, {Lane}, {Langgin}, {Lavender}, {Law}, {Lazarz}, {Leung}, {Leung}, {Lewis}, {Li}, {Li}, {Lian}, {Liang}, {Lin}, {Lin}, {Lin}, {Lintott}, {Long}, {Longa-Pe{\~n}a}, {L{\'o}pez-Cob{\'a}}, {Lu},
  {Lundgren}, {Luo}, {Mackereth}, {de la Macorra}, {Mahadevan}, {Majewski}, {Manchado}, {Mandeville}, {Maraston}, {Margalef-Bentabol}, {Masseron}, {Masters}, {Mathur}, {McDermid}, {Mckay}, {Merloni}, {Merrifield}, {Meszaros}, {Miglio}, {Di Mille}, {Minniti}, {Minsley}, \& {Monachesi}}]{Abdurro}
{Abdurro'uf}, {Accetta}, K., {Aerts}, C., {et~al.} 2022, \apjs, 259, 35

\bibitem[{{Alonso-Garc{\'\i}a} {et~al.}(2012){Alonso-Garc{\'\i}a}, {Mateo}, {Sen}, {Banerjee}, {Catelan}, {Minniti}, \& {von Braun}}]{AlonsoG2012}
{Alonso-Garc{\'\i}a}, J., {Mateo}, M., {Sen}, B., {et~al.} 2012, \aj, 143, 70

\bibitem[{{Aoki} {et~al.}(2007){Aoki}, {Beers}, {Christlieb}, {Norris}, {Ryan}, \& {Tsangarides}}]{Aoki2007}
{Aoki}, W., {Beers}, T.~C., {Christlieb}, N., {et~al.} 2007, \apj, 655, 492

\bibitem[{{Asplund} {et~al.}(2009){Asplund}, {Grevesse}, {Sauval}, \& {Scott}}]{Asplund09}
{Asplund}, M., {Grevesse}, N., {Sauval}, A.~J., \& {Scott}, P. 2009, \araa, 47, 481

\bibitem[{{Baumgardt} \& {Vasiliev}(2021)}]{BV2021}
{Baumgardt}, H., \& {Vasiliev}, E. 2021, \mnras, 505, 5957

\bibitem[{{Beers} \& {Christlieb}(2005)}]{BeersNC2005}
{Beers}, T.~C., \& {Christlieb}, N. 2005, \araa, 43, 531

\bibitem[{{Bessell} {et~al.}(2015){Bessell}, {Collet}, {Keller}, {Frebel}, {Heger}, {Casey}, {Masseron}, {Asplund}, {Jacobson}, {Lind}, {Marino}, {Norris}, {Yong}, {Da Costa}, {Chan}, {Magic}, {Schmidt}, \& {Tisserand}}]{MSB2015}
{Bessell}, M.~S., {Collet}, R., {Keller}, S.~C., {et~al.} 2015, \apjl, 806, L16

\bibitem[{{Bond}(1974)}]{Bond74}
{Bond}, H.~E. 1974, \apj, 194, 95

\bibitem[{{Briley} {et~al.}(1993){Briley}, {Smith}, {Hesser}, \& {Bell}}]{Briley1993}
{Briley}, M.~M., {Smith}, G.~H., {Hesser}, J.~E., \& {Bell}, R.~A. 1993, \aj, 106, 142

\bibitem[{{Carretta} {et~al.}(2009{\natexlab{a}}){Carretta}, {Bragaglia}, {Gratton}, {D'Orazi}, \& {Lucatello}}]{Carretta2009}
{Carretta}, E., {Bragaglia}, A., {Gratton}, R., {D'Orazi}, V., \& {Lucatello}, S. 2009{\natexlab{a}}, \aap, 508, 695

\bibitem[{{Carretta} {et~al.}(2009{\natexlab{b}}){Carretta}, {Bragaglia}, {Gratton}, \& {Lucatello}}]{CarrettaVIII}
{Carretta}, E., {Bragaglia}, A., {Gratton}, R., \& {Lucatello}, S. 2009{\natexlab{b}}, \aap, 505, 139

\bibitem[{{Carretta} {et~al.}(2009{\natexlab{c}}){Carretta}, {Bragaglia}, {Gratton}, {Lucatello}, {Catanzaro}, {Leone}, {Bellazzini}, {Claudi}, {D'Orazi}, {Momany}, {Ortolani}, {Pancino}, {Piotto}, {Recio-Blanco}, \& {Sabbi}}]{CarrettaVII}
{Carretta}, E., {Bragaglia}, A., {Gratton}, R.~G., {et~al.} 2009{\natexlab{c}}, \aap, 505, 117

\bibitem[{{Casagrande} \& {VandenBerg}(2018)}]{LC2018}
{Casagrande}, L., \& {VandenBerg}, D.~A. 2018, \mnras, 479, L102

\bibitem[{{Casagrande} {et~al.}(2021){Casagrande}, {Lin}, {Rains}, {Liu}, {Buder}, {Horner}, {Asplund}, {Lewis}, {Martell}, {Nordlander}, {Stello}, {Ting}, {Wittenmyer}, {Bland-Hawthorn}, {Casey}, {De Silva}, {D'Orazi}, {Freeman}, {Hayden}, {Kos}, {Lind}, {Schlesinger}, {Sharma}, {Simpson}, {Zucker}, \& {Zwitter}}]{Luca2021}
{Casagrande}, L., {Lin}, J., {Rains}, A.~D., {et~al.} 2021, \mnras, 507, 2684

\bibitem[{{Cordero} {et~al.}(2015){Cordero}, {Hansen}, {Johnson}, \& {Pilachowski}}]{Cordero15}
{Cordero}, M.~J., {Hansen}, C.~J., {Johnson}, C.~I., \& {Pilachowski}, C.~A. 2015, \apjl, 808, L10

\bibitem[{{C{\^o}t{\'e}} {et~al.}(1997){C{\^o}t{\'e}}, {Hanes}, {McLaughlin}, {Bridges}, {Hesser}, \& {Harris}}]{Cote1997}
{C{\^o}t{\'e}}, P., {Hanes}, D.~A., {McLaughlin}, D.~E., {et~al.} 1997, \apjl, 476, L15

\bibitem[{{Cristallo} {et~al.}(2009){Cristallo}, {Straniero}, {Gallino}, {Piersanti}, {Dom{\'\i}nguez}, \& {Lederer}}]{Cristallo09}
{Cristallo}, S., {Straniero}, O., {Gallino}, R., {et~al.} 2009, \apj, 696, 797

\bibitem[{{Da Costa} {et~al.}(2004){Da Costa}, {Cannon}, {Croke}, \& {Norris}}]{GDaC2004}
{Da Costa}, G.~S., {Cannon}, R., {Croke}, B., \& {Norris}, J. 2004, \memsai, 75, 370

\bibitem[{{Dahn} {et~al.}(1977){Dahn}, {Liebert}, {Kron}, {Spinrad}, \& {Hintzen}}]{Dahn77}
{Dahn}, C.~C., {Liebert}, J., {Kron}, R.~G., {Spinrad}, H., \& {Hintzen}, P.~M. 1977, \apj, 216, 757

\bibitem[{{de Jong} {et~al.}(2019){de Jong}, {Agertz}, {Berbel}, {Aird}, {Alexander}, {Amarsi}, {Anders}, {Andrae}, {Ansarinejad}, {Ansorge}, {Antilogus}, {Anwand-Heerwart}, {Arentsen}, {Arnadottir}, {Asplund}, {Auger}, {Azais}, {Baade}, {Baker}, {Baker}, {Balbinot}, {Baldry}, {Banerji}, {Barden}, {Barklem}, {Barth{\'e}l{\'e}my-Mazot}, {Battistini}, {Bauer}, {Bell}, {Bellido-Tirado}, {Bellstedt}, {Belokurov}, {Bensby}, {Bergemann}, {Bestenlehner}, {Bielby}, {Bilicki}, {Blake}, {Bland-Hawthorn}, {Boeche}, {Boland}, {Boller}, {Bongard}, {Bongiorno}, {Bonifacio}, {Boudon}, {Brooks}, {Brown}, {Brown}, {Br{\"u}ggen}, {Brynnel}, {Brzeski}, {Buchert}, {Buschkamp}, {Caffau}, {Caillier}, {Carrick}, {Casagrande}, {Case}, {Casey}, {Cesarini}, {Cescutti}, {Chapuis}, {Chiappini}, {Childress}, {Christlieb}, {Church}, {Cioni}, {Cluver}, {Colless}, {Collett}, {Comparat}, {Cooper}, {Couch}, {Courbin}, {Croom}, {Croton}, {Daguis{\'e}}, {Dalton}, {Davies}, {Davis}, {de Laverny}, {Deason}, {Dionies}, {Disseau}, {Doel},
  {D{\"o}scher}, {Driver}, {Dwelly}, {Eckert}, {Edge}, {Edvardsson}, {Youssoufi}, {Elhaddad}, {Enke}, {Erfanianfar}, {Farrell}, {Fechner}, {Feiz}, {Feltzing}, {Ferreras}, {Feuerstein}, {Feuillet}, {Finoguenov}, {Ford}, {Fotopoulou}, {Fouesneau}, {Frenk}, {Frey}, {Gaessler}, {Geier}, {Gentile Fusillo}, {Gerhard}, {Giannantonio}, {Giannone}, {Gibson}, {Gillingham}, {Gonz{\'a}lez-Fern{\'a}ndez}, {Gonzalez-Solares}, {Gottloeber}, {Gould}, {Grebel}, {Gueguen}, {Guiglion}, {Haehnelt}, {Hahn}, {Hansen}, {Hartman}, {Hauptner}, {Hawkins}, {Haynes}, {Haynes}, {Heiter}, {Helmi}, {Aguayo}, {Hewett}, {Hinton}, {Hobbs}, {Hoenig}, {Hofman}, {Hook}, {Hopgood}, {Hopkins}, {Hourihane}, {Howes}, {Howlett}, {Huet}, {Irwin}, {Iwert}, {Jablonka}, {Jahn}, {Jahnke}, {Jarno}, {Jin}, {Jofre}, {Johl}, {Jones}, {J{\"o}nsson}, {Jordan}, {Karovicova}, {Khalatyan}, {Kelz}, {Kennicutt}, {King}, {Kitaura}, {Klar}, {Klauser}, {Kneib}, {Koch}, {Koposov}, {Kordopatis}, {Korn}, {Kosmalski}, {Kotak}, {Kovalev}, {Kreckel}, {Kripak}, {Krumpe},
  {Kuijken}, {Kunder}, {Kushniruk}, {Lam}, {Lamer}, {Laurent}, {Lawrence}, {Lehmitz}, {Lemasle}, {Lewis}, {Li}, {Lidman}, {Lind}, {Liske}, {Lizon}, {Loveday}, {Ludwig}, {McDermid}, {Maguire}, {Mainieri}, {Mali}, \& {Mandel}}]{deJong2019}
{de Jong}, R.~S., {Agertz}, O., {Berbel}, A.~A., {et~al.} 2019, The Messenger, 175, 3

\bibitem[{{Dobrovolskas} {et~al.}(2021){Dobrovolskas}, {Kolomiecas}, {Ku{\v{c}}inskas}, {Klevas}, \& {Korotin}}]{Dobrov}
{Dobrovolskas}, V., {Kolomiecas}, E., {Ku{\v{c}}inskas}, A., {Klevas}, J., \& {Korotin}, S. 2021, \aap, 656, A67

\bibitem[{{D'Orazi} {et~al.}(2010){D'Orazi}, {Gratton}, {Lucatello}, {Carretta}, {Bragaglia}, \& {Marino}}]{DOrazi2010}
{D'Orazi}, V., {Gratton}, R., {Lucatello}, S., {et~al.} 2010, \apjl, 719, L213

\bibitem[{{Dotter} {et~al.}(2008){Dotter}, {Chaboyer}, {Jevremovi{\'c}}, {Kostov}, {Baron}, \& {Ferguson}}]{Dotter08}
{Dotter}, A., {Chaboyer}, B., {Jevremovi{\'c}}, D., {et~al.} 2008, \apjs, 178, 89

\bibitem[{{Foster} {et~al.}(2024){Foster}, {Schiavon}, {de Castro}, {Lucatello}, {Daher}, {Penoyre}, {Price-Whelan}, {Badenes}, {Fern{\'a}ndez-Trincado}, {Garc{\'\i}a-Hern{\'a}ndez}, {Holtzman}, {J{\"o}nsson}, \& {Shetrone}}]{Foster2024}
{Foster}, S., {Schiavon}, R.~P., {de Castro}, D.~B., {et~al.} 2024, \aap, 689, A230

\bibitem[{{Frebel} \& {Norris}(2015)}]{FN2015}
{Frebel}, A., \& {Norris}, J.~E. 2015, \araa, 53, 631

\bibitem[{{Gaia Collaboration} {et~al.}(2023){Gaia Collaboration}, {Vallenari}, {Brown}, {Prusti}, {de Bruijne}, {Arenou}, {Babusiaux}, {Biermann}, {Creevey}, {Ducourant}, {Evans}, {Eyer}, {Guerra}, {Hutton}, {Jordi}, {Klioner}, {Lammers}, {Lindegren}, {Luri}, {Mignard}, {Panem}, {Pourbaix}, {Randich}, {Sartoretti}, {Soubiran}, {Tanga}, {Walton}, {Bailer-Jones}, {Bastian}, {Drimmel}, {Jansen}, {Katz}, {Lattanzi}, {van Leeuwen}, {Bakker}, {Cacciari}, {Casta{\~n}eda}, {De Angeli}, {Fabricius}, {Fouesneau}, {Fr{\'e}mat}, {Galluccio}, {Guerrier}, {Heiter}, {Masana}, {Messineo}, {Mowlavi}, {Nicolas}, {Nienartowicz}, {Pailler}, {Panuzzo}, {Riclet}, {Roux}, {Seabroke}, {Sordo}, {Th{\'e}venin}, {Gracia-Abril}, {Portell}, {Teyssier}, {Altmann}, {Andrae}, {Audard}, {Bellas-Velidis}, {Benson}, {Berthier}, {Blomme}, {Burgess}, {Busonero}, {Busso}, {C{\'a}novas}, {Carry}, {Cellino}, {Cheek}, {Clementini}, {Damerdji}, {Davidson}, {de Teodoro}, {Nu{\~n}ez Campos}, {Delchambre}, {Dell'Oro}, {Esquej},
  {Fern{\'a}ndez-Hern{\'a}ndez}, {Fraile}, {Garabato}, {Garc{\'\i}a-Lario}, {Gosset}, {Haigron}, {Halbwachs}, {Hambly}, {Harrison}, {Hern{\'a}ndez}, {Hestroffer}, {Hodgkin}, {Holl}, {Jan{\ss}en}, {Jevardat de Fombelle}, {Jordan}, {Krone-Martins}, {Lanzafame}, {L{\"o}ffler}, {Marchal}, {Marrese}, {Moitinho}, {Muinonen}, {Osborne}, {Pancino}, {Pauwels}, {Recio-Blanco}, {Reyl{\'e}}, {Riello}, {Rimoldini}, {Roegiers}, {Rybizki}, {Sarro}, {Siopis}, {Smith}, {Sozzetti}, {Utrilla}, {van Leeuwen}, {Abbas}, {{\'A}brah{\'a}m}, {Abreu Aramburu}, {Aerts}, {Aguado}, {Ajaj}, {Aldea-Montero}, {Altavilla}, {{\'A}lvarez}, {Alves}, {Anders}, {Anderson}, {Anglada Varela}, {Antoja}, {Baines}, {Baker}, {Balaguer-N{\'u}{\~n}ez}, {Balbinot}, {Balog}, {Barache}, {Barbato}, {Barros}, {Barstow}, {Bartolom{\'e}}, {Bassilana}, {Bauchet}, {Becciani}, {Bellazzini}, {Berihuete}, {Bernet}, {Bertone}, {Bianchi}, {Binnenfeld}, {Blanco-Cuaresma}, {Blazere}, {Boch}, {Bombrun}, {Bossini}, {Bouquillon}, {Bragaglia}, {Bramante}, {Breedt},
  {Bressan}, {Brouillet}, {Brugaletta}, {Bucciarelli}, {Burlacu}, {Butkevich}, {Buzzi}, {Caffau}, {Cancelliere}, {Cantat-Gaudin}, {Carballo}, {Carlucci}, {Carnerero}, {Carrasco}, {Casamiquela}, {Castellani}, {Castro-Ginard}, {Chaoul}, {Charlot}, {Chemin}, {Chiaramida}, {Chiavassa}, {Chornay}, {Comoretto}, {Contursi}, {Cooper}, {Cornez}, {Cowell}, {Crifo}, {Cropper}, {Crosta}, {Crowley}, {Dafonte}, {Dapergolas}, {David}, {David}, {de Laverny}, {De Luise}, {De March}, {De Ridder}, {de Souza}, {de Torres}, {del Peloso}, {del Pozo}, {Delbo}, {Delgado}, {Delisle}, {Demouchy}, {Dharmawardena}, {Di Matteo}, {Diakite}, {Diener}, {Distefano}, {Dolding}, {Edvardsson}, {Enke}, {Fabre}, {Fabrizio}, {Faigler}, {Fedorets}, {Fernique}, {Fienga}, {Figueras}, {Fournier}, {Fouron}, {Fragkoudi}, {Gai}, {Garcia-Gutierrez}, {Garcia-Reinaldos}, {Garc{\'\i}a-Torres}, {Garofalo}, {Gavel}, {Gavras}, {Gerlach}, {Geyer}, {Giacobbe}, {Gilmore}, {Girona}, {Giuffrida}, {Gomel}, {Gomez}, {Gonz{\'a}lez-N{\'u}{\~n}ez},
  {Gonz{\'a}lez-Santamar{\'\i}a}, {Gonz{\'a}lez-Vidal}, {Granvik}, {Guillout}, {Guiraud}, {Guti{\'e}rrez-S{\'a}nchez}, {Guy}, {Hatzidimitriou}, {Hauser}, {Haywood}, {Helmer}, {Helmi}, {Sarmiento}, {Hidalgo}, {Hilger}, {H{\l}adczuk}, {Hobbs}, {Holland}, {Huckle}, {Jardine}, {Jasniewicz}, {Jean-Antoine Piccolo}, {Jim{\'e}nez-Arranz}, {Jorissen}, {Juaristi Campillo}, {Julbe}, {Karbevska}, {Kervella}, {Khanna}, {Kontizas}, {Kordopatis}, {Korn}, {K{\'o}sp{\'a}l}, {Kostrzewa-Rutkowska}, {Kruszy{\'n}ska}, {Kun}, {Laizeau}, {Lambert}, {Lanza}, {Lasne}, {Le Campion}, {Lebreton}, {Lebzelter}, {Leccia}, {Leclerc}, {Lecoeur-Taibi}, {Liao}, {Licata}, {Lindstr{\o}m}, {Lister}, {Livanou}, {Lobel}, {Lorca}, {Loup}, {Madrero Pardo}, {Magdaleno Romeo}, {Managau}, {Mann}, {Manteiga}, {Marchant}, {Marconi}, {Marcos}, {Marcos Santos}, {Mar{\'\i}n Pina}, {Marinoni}, {Marocco}, {Marshall}, {Martin Polo}, {Mart{\'\i}n-Fleitas}, {Marton}, {Mary}, {Masip}, {Massari}, {Mastrobuono-Battisti}, {Mazeh}, {McMillan}, {Messina}, {Michalik},
  {Millar}, {Mints}, {Molina}, {Molinaro}, {Moln{\'a}r}, {Monari}, {Mongui{\'o}}, {Montegriffo}, {Montero}, {Mor}, {Mora}, {Morbidelli}, {Morel}, {Morris}, {Muraveva}, {Murphy}, {Musella}, {Nagy}, {Noval}, {Oca{\~n}a}, {Ogden}, {Ordenovic}, {Osinde}, {Pagani}, {Pagano}, {Palaversa}, {Palicio}, {Pallas-Quintela}, {Panahi}, {Payne-Wardenaar}, {Pe{\~n}alosa Esteller}, {Penttil{\"a}}, {Pichon}, {Piersimoni}, {Pineau}, {Plachy}, {Plum}, {Poggio}, {Pr{\v{s}}a}, {Pulone}, {Racero}, {Ragaini}, {Rainer}, {Raiteri}, {Rambaux}, {Ramos}, {Ramos-Lerate}, {Re Fiorentin}, {Regibo}, {Richards}, {Rios Diaz}, {Ripepi}, {Riva}, {Rix}, {Rixon}, {Robichon}, {Robin}, {Robin}, {Roelens}, {Rogues}, {Rohrbasser}, {Romero-G{\'o}mez}, {Rowell}, {Royer}, {Ruz Mieres}, {Rybicki}, {Sadowski}, {S{\'a}ez N{\'u}{\~n}ez}, {Sagrist{\`a} Sell{\'e}s}, {Sahlmann}, {Salguero}, {Samaras}, {Sanchez Gimenez}, {Sanna}, {Santove{\~n}a}, {Sarasso}, {Schultheis}, {Sciacca}, {Segol}, {Segovia}, {S{\'e}gransan}, {Semeux}, {Shahaf}, {Siddiqui}, {Siebert},
  {Siltala}, {Silvelo}, {Slezak}, {Slezak}, {Smart}, {Snaith}, {Solano}, {Solitro}, {Souami}, {Souchay}, {Spagna}, {Spina}, {Spoto}, {Steele}, {Steidelm{\"u}ller}, {Stephenson}, {S{\"u}veges}, {Surdej}, {Szabados}, {Szegedi-Elek}, {Taris}, {Taylor}, {Teixeira}, {Tolomei}, {Tonello}, {Torra}, {Torra}, {Torralba Elipe}, {Trabucchi}, {Tsounis}, {Turon}, {Ulla}, {Unger}, {Vaillant}, {van Dillen}, {van Reeven}, {Vanel}, {Vecchiato}, {Viala}, {Vicente}, {Voutsinas}, {Weiler}, {Wevers}, {Wyrzykowski}, {Yoldas}, {Yvard}, {Zhao}, {Zorec}, {Zucker}, \& {Zwitter}}]{GaiaDR3}
{Gaia Collaboration}, {Vallenari}, A., {Brown}, A.~G.~A., {et~al.} 2023, \aap, 674, A1

\bibitem[{{Gerber} {et~al.}(2018){Gerber}, {Friel}, \& {Vesperini}}]{Gerber2018}
{Gerber}, J.~M., {Friel}, E.~D., \& {Vesperini}, E. 2018, \aj, 156, 6

\bibitem[{{Gerber} {et~al.}(2021){Gerber}, {Friel}, \& {Vesperini}}]{GFV2021}
---. 2021, \aj, 161, 288

\bibitem[{{Gonz{\'a}lez Hern{\'a}ndez} {et~al.}(2020){Gonz{\'a}lez Hern{\'a}ndez}, {Aguado}, {Allende Prieto}, {Burgasser}, \& {Rebolo}}]{GH2020}
{Gonz{\'a}lez Hern{\'a}ndez}, J.~I., {Aguado}, D.~S., {Allende Prieto}, C., {Burgasser}, A.~J., \& {Rebolo}, R. 2020, \apjl, 889, L13

\bibitem[{{Gustafsson} {et~al.}(2008){Gustafsson}, {Edvardsson}, {Eriksson}, {J{\o}rgensen}, {Nordlund}, \& {Plez}}]{Gustafsson2008}
{Gustafsson}, B., {Edvardsson}, B., {Eriksson}, K., {et~al.} 2008, \aap, 486, 951

\bibitem[{{Hesser} \& {Harris}(1979)}]{HesserH79}
{Hesser}, J.~E., \& {Harris}, G.~L.~H. 1979, \apj, 234, 513

\bibitem[{{Hut} {et~al.}(1992){Hut}, {McMillan}, {Goodman}, {Mateo}, {Phinney}, {Pryor}, {Richer}, {Verbunt}, \& {Weinberg}}]{Hut1992}
{Hut}, P., {McMillan}, S., {Goodman}, J., {et~al.} 1992, \pasp, 104, 981

\bibitem[{{Izzard} {et~al.}(2007){Izzard}, {Jeffery}, \& {Lattanzio}}]{Izzard2007}
{Izzard}, R.~G., {Jeffery}, C.~S., \& {Lattanzio}, J. 2007, \aap, 470, 661

\bibitem[{{Jin} {et~al.}(2024){Jin}, {Trager}, {Dalton}, {Aguerri}, {Drew}, {Falc{\'o}n-Barroso}, {G{\"a}nsicke}, {Hill}, {Iovino}, {Pieri}, {Poggianti}, {Smith}, {Vallenari}, {Abrams}, {Aguado}, {Antoja}, {Arag{\'o}n-Salamanca}, {Ascasibar}, {Babusiaux}, {Balcells}, {Barrena}, {Battaglia}, {Belokurov}, {Bensby}, {Bonifacio}, {Bragaglia}, {Carrasco}, {Carrera}, {Cornwell}, {Dom{\'\i}nguez-Palmero}, {Duncan}, {Famaey}, {Fari{\~n}a}, {Gonzalez}, {Guest}, {Hatch}, {Hess}, {Hoskin}, {Irwin}, {Knapen}, {Koposov}, {Kuchner}, {Laigle}, {Lewis}, {Longhetti}, {Lucatello}, {M{\'e}ndez-Abreu}, {Mercurio}, {Molaeinezhad}, {Mongui{\'o}}, {Morrison}, {Murphy}, {Peralta de Arriba}, {P{\'e}rez}, {P{\'e}rez-R{\`a}fols}, {Pic{\'o}}, {Raddi}, {Romero-G{\'o}mez}, {Royer}, {Siebert}, {Seabroke}, {Som}, {Terrett}, {Thomas}, {Wesson}, {Worley}, {Alfaro}, {Allende Prieto}, {Alonso-Santiago}, {Amos}, {Ashley}, {Balaguer-N{\'u}{\~n}ez}, {Balbinot}, {Bellazzini}, {Benn}, {Berlanas}, {Bernard}, {Best}, {Bettoni}, {Bianco}, {Bishop},
  {Blomqvist}, {Boeche}, {Bolzonella}, {Bonoli}, {Bosma}, {Britavskiy}, {Busarello}, {Caffau}, {Cantat-Gaudin}, {Castro-Ginard}, {Couto}, {Carbajo-Hijarrubia}, {Carter}, {Casamiquela}, {Conrado}, {Corcho-Caballero}, {Costantin}, {Deason}, {de Burgos}, {De Grandi}, {Di Matteo}, {Dom{\'\i}nguez-G{\'o}mez}, {Dorda}, {Drake}, {Dutta}, {Erkal}, {Feltzing}, {Ferr{\'e}-Mateu}, {Feuillet}, {Figueras}, {Fossati}, {Franciosini}, {Frasca}, {Fumagalli}, {Gallazzi}, {Garc{\'\i}a-Benito}, {Gentile Fusillo}, {Gebran}, {Gilbert}, {Gledhill}, {Gonz{\'a}lez Delgado}, {Greimel}, {Guarcello}, {Guerra}, {Gullieuszik}, {Haines}, {Hardcastle}, {Harris}, {Haywood}, {Helmi}, {Hernandez}, {Herrero}, {Hughes}, {Ir{\v{s}}i{\v{c}}}, {Jablonka}, {Jarvis}, {Jordi}, {Kondapally}, {Kordopatis}, {Krogager}, {La Barbera}, {Lam}, {Larsen}, {Lemasle}, {Lewis}, {Lhom{\'e}}, {Lind}, {Lodi}, {Longobardi}, {Lonoce}, {Magrini}, {Ma{\'\i}z Apell{\'a}niz}, {Marchal}, {Marco}, {Martin}, {Matsuno}, {Maurogordato}, {Merluzzi}, {Miralda-Escud{\'e}},
  {Molinari}, {Monari}, {Morelli}, {Mottram}, {Naylor}, {Negueruela}, {O{\~n}orbe}, {Pancino}, {Peirani}, {Peletier}, {Pozzetti}, {Rainer}, {Ramos}, {Read}, {Rossi}, {R{\"o}ttgering}, {Rubi{\~n}o-Mart{\'\i}n}, {Sabater}, {San Juan}, {Sanna}, {Schallig}, {Schiavon}, {Schultheis}, {Serra}, {Shimwell}, {Sim{\'o}n-D{\'\i}az}, {Smith}, {Sordo}, {Sorini}, {Soubiran}, {Starkenburg}, {Steele}, {Stott}, {Stuik}, {Tolstoy}, {Tortora}, {Tsantaki}, {Van der Swaelmen}, {van Weeren}, \& {Vergani}}]{Shoko2024}
{Jin}, S., {Trager}, S.~C., {Dalton}, G.~B., {et~al.} 2024, \mnras, 530, 2688

\bibitem[{{Karakas} \& {Lattanzio}(2007)}]{AKJL07}
{Karakas}, A., \& {Lattanzio}, J.~C. 2007, \pasa, 24, 103

\bibitem[{{Keenan}(1942)}]{Keenan42}
{Keenan}, P.~C. 1942, \apj, 96, 101

\bibitem[{{Kirby} {et~al.}(2015){Kirby}, {Guo}, {Zhang}, {Deng}, {Cohen}, {Guhathakurta}, {Shetrone}, {Lee}, \& {Rizzi}}]{Kirby2015}
{Kirby}, E.~N., {Guo}, M., {Zhang}, A.~J., {et~al.} 2015, \apj, 801, 125

\bibitem[{{Kremer} {et~al.}(2020){Kremer}, {Ye}, {Rui}, {Weatherford}, {Chatterjee}, {Fragione}, {Rodriguez}, {Spera}, \& {Rasio}}]{Kremer2020}
{Kremer}, K., {Ye}, C.~S., {Rui}, N.~Z., {et~al.} 2020, \apjs, 247, 48

\bibitem[{{Landolt}(1992)}]{Landolt92}
{Landolt}, A.~U. 1992, \aj, 104, 340

\bibitem[{{Lardo} {et~al.}(2016){Lardo}, {Battaglia}, {Pancino}, {Romano}, {de Boer}, {Starkenburg}, {Tolstoy}, {Irwin}, {Jablonka}, \& {Tosi}}]{Lardo2016}
{Lardo}, C., {Battaglia}, G., {Pancino}, E., {et~al.} 2016, \aap, 585, A70

\bibitem[{{Latour} {et~al.}(2025){Latour}, {Kamann}, {Martocchia}, {Husser}, {Saracino}, \& {Dreizler}}]{Latour25}
{Latour}, M., {Kamann}, S., {Martocchia}, S., {et~al.} 2025, \aap, 694, A248

\bibitem[{{Legnardi} {et~al.}(2022){Legnardi}, {Milone}, {Armillotta}, {Marino}, {Cordoni}, {Renzini}, {Vesperini}, {D'Antona}, {McKenzie}, {Yong}, {Dondoglio}, {Lagioia}, {Carlos}, {Tailo}, {Jang}, \& {Mohandasan}}]{Legnardi22}
{Legnardi}, M.~V., {Milone}, A.~P., {Armillotta}, L., {et~al.} 2022, \mnras, 513, 735

\bibitem[{{Lewis} {et~al.}(2002){Lewis}, {Cannon}, {Taylor}, {Glazebrook}, {Bailey}, {Baldry}, {Barton}, {Bridges}, {Dalton}, {Farrell}, {Gray}, {Lankshear}, {McCowage}, {Parry}, {Sharples}, {Shortridge}, {Smith}, {Stevenson}, {Straede}, {Waller}, {Whittard}, {Wilcox}, \& {Willis}}]{Lewis2002}
{Lewis}, I.~J., {Cannon}, R.~D., {Taylor}, K., {et~al.} 2002, \mnras, 333, 279

\bibitem[{{Li} {et~al.}(2024){Li}, {Zhang}, {Cui}, {Shi}, {Ji}, {Huo}, {Gao}, {Zhang}, \& {Sun}}]{Lin2024}
{Li}, L., {Zhang}, K., {Cui}, W., {et~al.} 2024, \apjs, 271, 12

\bibitem[{{Lucatello} {et~al.}(2005){Lucatello}, {Tsangarides}, {Beers}, {Carretta}, {Gratton}, \& {Ryan}}]{Lucatello2005}
{Lucatello}, S., {Tsangarides}, S., {Beers}, T.~C., {et~al.} 2005, \apj, 625, 825

\bibitem[{{Malavolta} {et~al.}(2014){Malavolta}, {Sneden}, {Piotto}, {Milone}, {Bedin}, \& {Nascimbeni}}]{Malavolta}
{Malavolta}, L., {Sneden}, C., {Piotto}, G., {et~al.} 2014, \aj, 147, 25

\bibitem[{{Masseron} {et~al.}(2014){Masseron}, {Plez}, {Van Eck}, {Colin}, {Daoutidis}, {Godefroid}, {Coheur}, {Bernath}, {Jorissen}, \& {Christlieb}}]{Masseron2014}
{Masseron}, T., {Plez}, B., {Van Eck}, S., {et~al.} 2014, \aap, 571, A47

\bibitem[{{McClure}(1984)}]{McClure1984}
{McClure}, R.~D. 1984, \apjl, 280, L31

\bibitem[{{McClure}(1997{\natexlab{a}})}]{McClure1997a}
---. 1997{\natexlab{a}}, \pasp, 109, 536

\bibitem[{{McClure}(1997{\natexlab{b}})}]{McClure1997b}
---. 1997{\natexlab{b}}, \pasp, 109, 256

\bibitem[{{McClure} \& {Norris}(1977)}]{McJEN1977}
{McClure}, R.~D., \& {Norris}, J. 1977, \apjl, 217, L101

\bibitem[{{McClure} \& {Woodsworth}(1990)}]{McClureWood}
{McClure}, R.~D., \& {Woodsworth}, A.~W. 1990, \apj, 352, 709

\bibitem[{{Milone} {et~al.}(2017){Milone}, {Piotto}, {Renzini}, {Marino}, {Bedin}, {Vesperini}, {D'Antona}, {Nardiello}, {Anderson}, {King}, {Yong}, {Bellini}, {Aparicio}, {Barbuy}, {Brown}, {Cassisi}, {Ortolani}, {Salaris}, {Sarajedini}, \& {van der Marel}}]{Milone17}
{Milone}, A.~P., {Piotto}, G., {Renzini}, A., {et~al.} 2017, \mnras, 464, 3636

\bibitem[{{Monet}(1996)}]{Monet96}
{Monet}, D., ed. 1996, {USNO A - 1.0 a catalog of astrometric standards}

\bibitem[{{Nomoto} {et~al.}(2013){Nomoto}, {Kobayashi}, \& {Tominaga}}]{NKT2013}
{Nomoto}, K., {Kobayashi}, C., \& {Tominaga}, N. 2013, \araa, 51, 457

\bibitem[{{Nordlander} {et~al.}(2019){Nordlander}, {Bessell}, {Da Costa}, {Mackey}, {Asplund}, {Casey}, {Chiti}, {Ezzeddine}, {Frebel}, {Lind}, {Marino}, {Murphy}, {Norris}, {Schmidt}, \& {Yong}}]{TN2019}
{Nordlander}, T., {Bessell}, M.~S., {Da Costa}, G.~S., {et~al.} 2019, \mnras, 488, L109

\bibitem[{Ochsenbein(1996)}]{10.26093/cds/vizier}
Ochsenbein, F. 1996, The VizieR database of astronomical catalogues, doi:10.26093/CDS/VIZIER

\bibitem[{{Ochsenbein} {et~al.}(2000){Ochsenbein}, {Bauer}, \& {Marcout}}]{vizier2000}
{Ochsenbein}, F., {Bauer}, P., \& {Marcout}, J. 2000, \aaps, 143, 23

\bibitem[{{Pastorelli} {et~al.}(2020){Pastorelli}, {Marigo}, {Girardi}, {Aringer}, {Chen}, {Rubele}, {Trabucchi}, {Bladh}, {Boyer}, {Bressan}, {Dalcanton}, {Groenewegen}, {Lebzelter}, {Mowlavi}, {Chubb}, {Cioni}, {de Grijs}, {Ivanov}, {Nanni}, {van Loon}, \& {Zaggia}}]{Pasto2020}
{Pastorelli}, G., {Marigo}, P., {Girardi}, L., {et~al.} 2020, \mnras, 498, 3283

\bibitem[{{Placco} {et~al.}(2014){Placco}, {Frebel}, {Beers}, \& {Stancliffe}}]{Placco14}
{Placco}, V.~M., {Frebel}, A., {Beers}, T.~C., \& {Stancliffe}, R.~J. 2014, \apj, 797, 21

\bibitem[{{Plez}(2012)}]{Plez2012}
{Plez}, B. 2012, {Turbospectrum: Code for spectral synthesis}, Astrophysics Source Code Library, record ascl:1205.004

\bibitem[{{Rain} {et~al.}(2019){Rain}, {Villanova}, {Mun{\~o}z}, \& {Valenzuela-Calderon}}]{Rain2019}
{Rain}, M.~J., {Villanova}, S., {Mun{\~o}z}, C., \& {Valenzuela-Calderon}, C. 2019, \mnras, 483, 1674

\bibitem[{{Roederer} {et~al.}(2016){Roederer}, {Mateo}, {Bailey}, {Spencer}, {Crane}, \& {Shectman}}]{Roderer16}
{Roederer}, I.~U., {Mateo}, M., {Bailey}, J.~I., {et~al.} 2016, \mnras, 455, 2417

\bibitem[{{Rosenfield} {et~al.}(2016){Rosenfield}, {Marigo}, {Girardi}, {Dalcanton}, {Bressan}, {Williams}, \& {Dolphin}}]{Rosenfield16}
{Rosenfield}, P., {Marigo}, P., {Girardi}, L., {et~al.} 2016, \apj, 822, 73

\bibitem[{{Ryabchikova} \& {Pakhomov}(2015)}]{VLAD}
{Ryabchikova}, T., \& {Pakhomov}, Y. 2015, Baltic Astronomy, 24, 453

\bibitem[{{Salgado} {et~al.}(2016){Salgado}, {Da Costa}, {Yong}, \& {Norris}}]{Salgardo2016}
{Salgado}, C., {Da Costa}, G.~S., {Yong}, D., \& {Norris}, J.~E. 2016, \mnras, 463, 598

\bibitem[{{Sharina} {et~al.}(2012){Sharina}, {Aringer}, {Davoust}, {Kniazev}, \& {Donzelli}}]{Sharina2012}
{Sharina}, M., {Aringer}, B., {Davoust}, E., {Kniazev}, A.~Y., \& {Donzelli}, C.~J. 2012, \mnras, 426, L31

\bibitem[{{Smith} \& {Mateo}(1990)}]{SmithMateo1990}
{Smith}, G.~H., \& {Mateo}, M. 1990, \apj, 353, 533

\bibitem[{{Smith} \& {Norris}(1982)}]{GHSJEN1982}
{Smith}, G.~H., \& {Norris}, J. 1982, \apj, 254, 149

\bibitem[{{Stanford} {et~al.}(2006){Stanford}, {Da Costa}, {Norris}, \& {Cannon}}]{LS2006}
{Stanford}, L.~M., {Da Costa}, G.~S., {Norris}, J.~E., \& {Cannon}, R.~D. 2006, \apj, 647, 1075

\bibitem[{{Stanford} {et~al.}(2007){Stanford}, {Da Costa}, {Norris}, \& {Cannon}}]{LS2007}
---. 2007, \apj, 667, 911

\bibitem[{{Starkenburg} {et~al.}(2014){Starkenburg}, {Shetrone}, {McConnachie}, \& {Venn}}]{Starkenburg2014}
{Starkenburg}, E., {Shetrone}, M.~D., {McConnachie}, A.~W., \& {Venn}, K.~A. 2014, \mnras, 441, 1217

\bibitem[{{Usman} {et~al.}(2024){Usman}, {Ji}, {Li}, {Pace}, {Cullinane}, {Da Costa}, {Koposov}, {Lewis}, {Zucker}, {Belokurov}, {Bland-Hawthorn}, {Ferguson}, {Hansen}, {Limberg}, {Martell}, {McKenzie}, {Simon}, \& {S5 Collaboration}}]{Usman2024}
{Usman}, S.~A., {Ji}, A.~P., {Li}, T.~S., {et~al.} 2024, \mnras, 529, 2413

\bibitem[{{van Loon} {et~al.}(2007){van Loon}, {van Leeuwen}, {Smalley}, {Smith}, {Lyons}, {McDonald}, \& {Boyer}}]{vanLoon2007}
{van Loon}, J.~T., {van Leeuwen}, F., {Smalley}, B., {et~al.} 2007, \mnras, 382, 1353

\bibitem[{{Vasiliev} \& {Baumgardt}(2021)}]{VB2021}
{Vasiliev}, E., \& {Baumgardt}, H. 2021, \mnras, 505, 5978

\bibitem[{{Ventura} \& {D'Antona}(2009)}]{VA09}
{Ventura}, P., \& {D'Antona}, F. 2009, \aap, 499, 835

\bibitem[{{Vernet} {et~al.}(2011){Vernet}, {Dekker}, {D'Odorico}, {Kaper}, {Kjaergaard}, {Hammer}, {Randich}, {Zerbi}, {Groot}, {Hjorth}, {Guinouard}, {Navarro}, {Adolfse}, {Albers}, {Amans}, {Andersen}, {Andersen}, {Binetruy}, {Bristow}, {Castillo}, {Chemla}, {Christensen}, {Conconi}, {Conzelmann}, {Dam}, {de Caprio}, {de Ugarte Postigo}, {Delabre}, {di Marcantonio}, {Downing}, {Elswijk}, {Finger}, {Fischer}, {Flores}, {Fran{\c{c}}ois}, {Goldoni}, {Guglielmi}, {Haigron}, {Hanenburg}, {Hendriks}, {Horrobin}, {Horville}, {Jessen}, {Kerber}, {Kern}, {Kiekebusch}, {Kleszcz}, {Klougart}, {Kragt}, {Larsen}, {Lizon}, {Lucuix}, {Mainieri}, {Manuputy}, {Martayan}, {Mason}, {Mazzoleni}, {Michaelsen}, {Modigliani}, {Moehler}, {M{\o}ller}, {Norup S{\o}rensen}, {N{\o}rregaard}, {P{\'e}roux}, {Patat}, {Pena}, {Pragt}, {Reinero}, {Rigal}, {Riva}, {Roelfsema}, {Royer}, {Sacco}, {Santin}, {Schoenmaker}, {Spano}, {Sweers}, {Ter Horst}, {Tintori}, {Tromp}, {van Dael}, {van der Vliet}, {Venema}, {Vidali}, {Vinther}, {Vola},
  {Winters}, {Wistisen}, {Wulterkens}, \& {Zacchei}}]{vernet11}
{Vernet}, J., {Dekker}, H., {D'Odorico}, S., {et~al.} 2011, \aap, 536, A105

\bibitem[{{Wallace} \& {Gray}(2014)}]{Wallace2014}
{Wallace}, P.~T., \& {Gray}, N. 2014, {ASTROM: Basic astrometry program}, Astrophysics Source Code Library, record ascl:1406.008

\bibitem[{{Yong} {et~al.}(2014){Yong}, {Roederer}, {Grundahl}, {Da Costa}, {Karakas}, {Norris}, {Aoki}, {Fishlock}, {Marino}, {Milone}, \& {Shingles}}]{Yong2014}
{Yong}, D., {Roederer}, I.~U., {Grundahl}, F., {et~al.} 2014, \mnras, 441, 3396

\bibitem[{{Zhang} \& {Yuan}(2023)}]{ZhangYuan2023}
{Zhang}, R., \& {Yuan}, H. 2023, \apjs, 264, 14

\bibitem[{{Zhang} {et~al.}(2020){Zhang}, {Jeffery}, {Li}, \& {Bi}}]{Zhang2020}
{Zhang}, X., {Jeffery}, C.~S., {Li}, Y., \& {Bi}, S. 2020, \apj, 889, 33

\end{thebibliography}

\end{document}